\documentclass[aip,reprint,twocolumn,twoside]{revtex4-1}

\usepackage{graphicx}
\usepackage{dcolumn}
\usepackage{placeins}

\usepackage[version=3]{mhchem} 
\usepackage{amssymb}
\usepackage{multirow}
\usepackage{graphicx}
\usepackage{amsmath}
\usepackage{threeparttable}
\usepackage{float}
\usepackage{hyperref}
\usepackage{hyperref}
\usepackage{caption}
\usepackage{ragged2e}

\DeclareCaptionFormat{withunderline}{#1#2#3\par\vspace{1pt}\hrule\vspace{1pt}}

\usepackage{listings}
\usepackage{xcolor}
\usepackage{todonotes}
\usepackage{makecell}

\usepackage[utf8]{inputenc} 
\usepackage[T1]{fontenc}    

\newcommand{\paragraphtitle}[1]{\textsf{\textbf{\small {#1}}}}

\begin{document}

\author{Kai Zhu}
\thanks{Contributed equally to this work}
\affiliation{\small College of Pharmaceutical Sciences, Zhejiang University, Hangzhou, 310058, Zhejiang, China}
\affiliation{\small Atomistic Simulations, Italian Institute of Technology, Genova 16152, Italy}
\author{Jintu Zhang}
\thanks{Contributed equally to this work}
\affiliation{\small College of Pharmaceutical Sciences, Zhejiang University, Hangzhou, 310058, Zhejiang, China}
\author{Pietro Novelli}
\affiliation{\small Computational Statistics and Machine Learning, Italian Institute of Technology, Genova 16152, Italy}
\author{Tingjun Hou}
\email{tingjunhou@zju.edu.cn}
\affiliation{\small College of Pharmaceutical Sciences, Zhejiang University, Hangzhou, 310058, Zhejiang, China}
\author{Luigi Bonati}
\email{luigi.bonati@iit.it}
\affiliation{\small Atomistic Simulations, Italian Institute of Technology, Genova 16152, Italy}

\title{SelfTICA: contrastive learning of dynamical representations for rare-event sampling and characterization}

\begin{abstract}
Rare events govern many important molecular processes but remain difficult to characterize within accessible simulation timescales. This challenge has motivated machine-learning methods to construct low-dimensional collective variables for specific objectives, including state discrimination, slow-mode identification, and committor-function approximation. Here we introduce SelfTICA, a self-supervised framework that uses contrastive learning on time-lagged configurations to learn a latent representation of the slow modes governing relevant transitions. Once learned, this representation is frozen and reused across downstream tasks, including collective variables construction, enhanced sampling, free-energy estimation, and committor learning. Compared with direct slow-mode optimization, SelfTICA improves training stability and provides collective variables even from limited and exploratory trajectories, accelerating rare-event sampling and enabling accurate free-energy estimation. The dynamical information encoded during pretraining also accelerates committor learning and enables characterization of transition-state regions. These results show that contrastive learning of slow dynamical representations provides a common foundation for rare-event sampling and mechanistic analysis.
\end{abstract}

\maketitle

\twocolumngrid

\section{INTRODUCTION}
\label{sec:intro}

Atomistic simulations provide detailed insight into microscopic processes in physics, chemistry, and biology \cite{frenkel2023understanding}, but their practical reach is often limited by rare events occurring on timescales beyond those routinely accessible to conventional atomistic simulations. Enhanced sampling methods address this limitation by accelerating configurational exploration through bias potentials acting on low-dimensional functions of molecular configurations, commonly referred to as collective variables (CVs) \cite{henin2022enhanced,valsson2016enhancing}. 
In general, CVs are constructed from molecular descriptions ranging from hand-crafted geometric descriptors to representations learned directly from simulation data. For enhanced sampling applications, these descriptions must retain the information needed to resolve the slow molecular evolution of the system. The central challenge is therefore not only to reduce the dimensionality of molecular configurations, but also to identify features whose temporal evolution reflects the processes of interest.

Data-driven methods provide a natural route to identifying such features and have increasingly been used to construct molecular representations and low-dimensional coordinates\cite{mehdi2024enhanced,zhu2025enhanced}. Supervised classifiers learn coordinates that discriminate predefined metastable states \cite{bonati2020data,trizio2021enhanced}, whereas unsupervised approaches such as autoencoders seek embeddings from which molecular configurations can be reconstructed\cite{chen2018molecular,chen2018collective,belkacemi2021chasing,wang2019past,wang2021state}. Although these methods can provide useful structural representations, their objectives do not explicitly target the temporal evolution. Molecular kinetics is addressed more directly by state-specific quantities such as the committor function~\cite{vanden2010transition,jung2023machine,kang2024computing,trizio2025everything,megias2025iterative} or by spectral decomposition of dynamical operators~\cite{prinz2011markov,perez2013identification}. For a selected pair of metastable states $A$ and $B$, the committor assigns to each configuration the probability of reaching $B$ before $A$, thereby providing a natural reaction coordinate for the transition between the two states~\cite{vanden2010transition}. Spectral approaches instead characterize global slow dynamics through the dominant modes of the transfer or Koopman operator. These include the variational approach for conformational dynamics (VAC) and the variational approach for Markov processes (VAMP), encompassing linear methods such as TICA as well as neural-network approaches for learning nonlinear slow coordinates~\cite{prinz2011markov,perez2013identification,nuske2014variational,wu2017variational,molgedey1994separation,bonati2021deep,chen2019nonlinear,shmilovich2023girsanov,mardt2018vampnets}.

\begin{figure*}[t!]  
\centering
\includegraphics[width=1.00\linewidth]{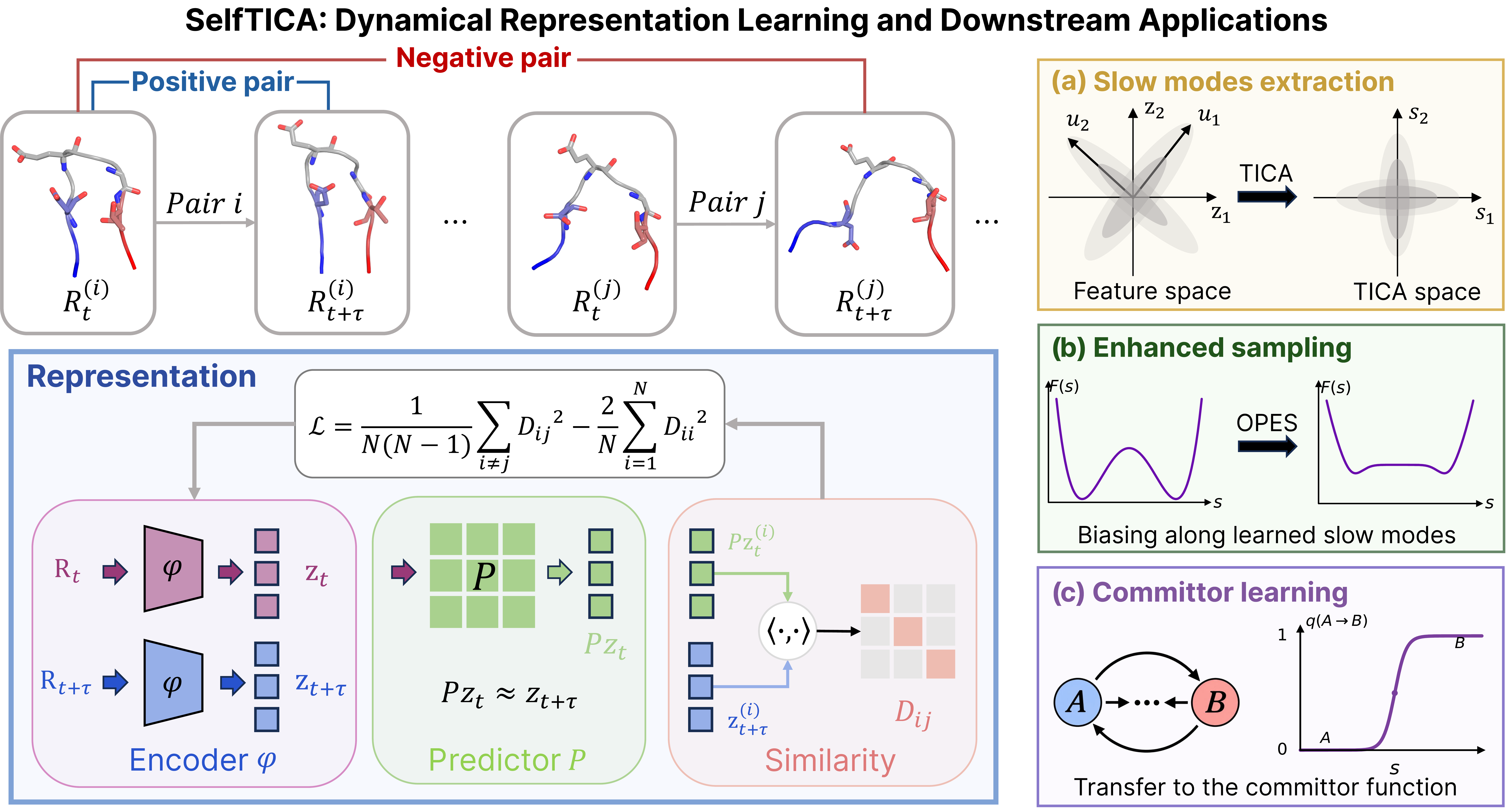}
\caption{\justifying \textbf{SelfTICA: dynamical representation learning and downstream applications.}
(Top left) Time-lagged configurations from the same trajectory pair define positive examples, whereas mismatched configurations define negative examples.
(Bottom left) A shared encoder $\varphi$ maps configurations to latent features, and a predictor $P$ propagates the representation forward in time. Pairwise similarities $D_{ij}$ between predicted and target features define the contrastive objective in Eq.~\ref{eq:em_loss}.
(Right) The learned representation supports multiple downstream tasks: (a) extraction of slow modes by TICA, (b) enhanced sampling using the learned slow modes as CVs, and (c) transfer to committor learning through a task-specific readout.}
\label{fig:method}
\end{figure*}

Although these methods target different aspects of molecular kinetics, the learned features are generally tied to the specific observable being estimated. Indeed, the committor is defined for a selected pair of states, and learning it requires the boundary states to be specified in advance. Furthermore, accurate committor estimation requires data from low-probability transition regions, which in turn often entails repeated cycles of model optimization and sampling~\cite{jung2023machine,kang2024computing,trizio2025everything,megias2025iterative}. Neural variational methods such as DeepTICA~\cite{bonati2021deep} and state-free reversible VAMPnets (SRVs)~\cite{chen2019nonlinear} instead learn slow modes by directly optimizing spectral quantities at a prescribed lag time, thus coupling the learned features to the selected analysis timescale. The lag time must therefore be chosen \textit{a priori}, and changing it may require retraining, while differentiation through eigendecompositions can become numerically delicate near spectral degeneracies~\cite{wang2019backpropagation,wang2021robust}. These approaches further rely on time-lagged data that faithfully reflect the underlying dynamics, a requirement that is difficult to satisfy for complex systems where equilibrium trajectories containing sufficient transitions are rarely available. Training data are therefore often obtained from biased exploratory simulations using generic or suboptimal CVs~\cite{yang2018refining,bonati2021deep,shmilovich2023girsanov,devergne2024biased,devergne2025slow}, where reweighting is particularly challenging. 

These considerations motivate a different strategy focused on the learning of a dynamical representation. Rather than learning a new nonlinear coordinate for each CV objective, one could first learn a representation that retains broadly useful kinetic information and subsequently use it as a common feature basis for different downstream tasks. Self-supervised pretraining provides a natural route toward this goal because it can extract reusable features directly from unlabeled trajectory data~\cite{ross2022large,fang2022geometry,zhou2023uni}.  Among self-supervised approaches, contrastive learning builds representations by comparing pairs of matched and mismatched samples. This has been particularly successful in computer vision \cite{chen2020simple,he2020momentum,radford2021learning} and has also been applied to molecular systems \cite{zeng2022accurate,wang2022molecular}. The way matched pairs are constructed determines which statistical relationships the representation is encouraged to preserve. Whereas matched pairs commonly correspond to different views or augmentations of the same object, pairing configurations separated in time allows us to focus the content of the representation on temporal correlations.

Based on this idea, we introduce SelfTICA, a self-supervised framework for learning dynamical representations from time-lagged molecular configurations (see Fig.~\ref{fig:method} for a graphical overview). For a given molecular system, the nonlinear representation is learned once and subsequently kept fixed, while different kinetic objectives are addressed through inexpensive post-training operations or by adding lightweight task-specific readouts. 
During training, we sample time-lagged configuration pairs $(\mathbf{R}_t^{(i)}, \mathbf{R}_{t+\tau}^{(i)})$ from the same molecular trajectory. Configurations from the same sampled pair are treated as matched, or positive, pairs, whereas configurations from different sampled pairs, $(\mathbf{R}_t^{(i)}, \mathbf{R}_{t+\tau}^{(j)})$ with $i \neq j$, are treated as mismatched, or negative, pairs. A shared encoder $\varphi$ maps the configurations to latent features, and a predictor $P$ estimates the future representation $\varphi(\mathbf{R}_{t+\tau})$ from the initial one $\varphi(\mathbf{R}_t)$. The encoder and predictor are jointly optimized to increase the similarity between predicted and target features for matched pairs while decreasing it for mismatched pairs. Intuitively, this procedure seeks a latent space in which evolution over the lag time can be captured by a lightweight predictor. Notably, the resulting contrastive objective is connected to the VAMP-2 variational principle \cite{turri2026selfsupervised}, thereby favoring representations enriched in slow dynamical information without requiring differentiation through an eigenvalue problem during training.

Once the representation has been learned, it can be used to address several central problems in molecular sampling. TICA can be applied \textit{a posteriori} to the frozen representation at different lag times, without retraining the encoder, to extract orthogonal slow dynamical modes. These modes reveal the dominant long-timescale processes and provide natural coordinates for identifying metastable states. When used as CVs for enhanced sampling, they accelerate transitions between such states and extend the timescales accessible to simulations, enabling accurate free-energy calculations. The same representation can be used to estimate state-specific quantities such as the committor function, which is a central quantity in molecular simulations since it provides a measure of the progress of the reaction and enables the identification and characterization of transition-state ensembles. 

We demonstrate these capabilities through a sequence of molecular simulation tasks. We first show that SelfTICA learns representations that recover dominant slow modes, decouple nonlinear representation learning from the choice of analysis timescale, and improve the stability of slow-mode learning compared with direct spectral optimization. We then show that these representations yield effective CVs for enhanced sampling from limited and exploratory trajectories. By combining SelfTICA with graph neural network (GNN) encoders, we extend the framework to dynamically changing solvent and surface environments. Finally, we transfer frozen SelfTICA representations to committor learning, substantially reducing the iterative sampling and refinement required for transition-state characterization. These results show that SelfTICA provides a common nonlinear feature basis (representation) for studying global slow dynamics, accelerating rare-event sampling, and resolving transition mechanisms.

\begin{figure*}[t] 
\centering
\includegraphics[width=1.00\linewidth]{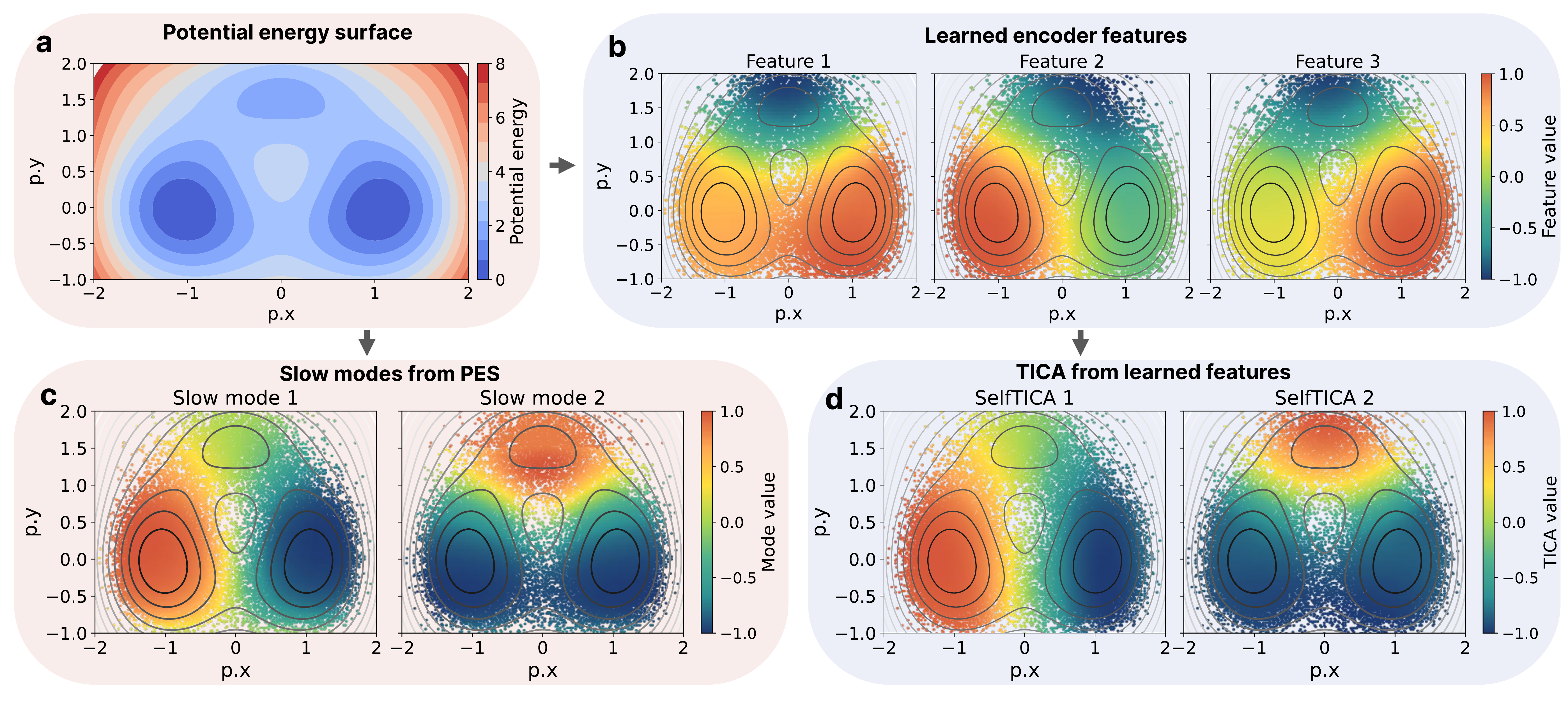}
\caption{\justifying \textbf{SelfTICA recovers the dominant slow modes of the triple-well system.}
(a) Potential energy surface in the particle coordinates $(p_x,p_y)$.
(b) Representative features learned by the SelfTICA encoder from an unbiased trajectory.
(c) Reference slow modes obtained from the spectral decomposition of the stochastic generator.
(d) TICA components extracted from the learned feature space, which closely reproduce the two reference slow modes in (c).
}
\label{fig:slow-modes}
\end{figure*}

\section{Results}

\subsection{SelfTICA learns slow dynamical representations}

We first show that our method learns a feature space that captures the dominant slow dynamics,  using an analytical toy model (Fig.~\ref{fig:slow-modes}a) for which reference slow modes can be obtained from the spectral decomposition of the underlying stochastic generator (Fig.~\ref{fig:slow-modes}c). SelfTICA was trained on unbiased trajectories generated at $k_{\mathrm{B}}T=1.0$ (Supplementary Fig.~1), where spontaneous barrier crossings provide sufficient information about the dominant interbasin dynamics. Fig.~\ref{fig:slow-modes}b shows that the learned encoder features resolve the principal metastable states and capture the slowest dynamical modes connected to transitions between states, but they are not necessarily orthogonal or ordered by relaxation timescale. Applying TICA in the learned feature space yields orthogonal slow modes that closely reproduce the reference eigenfunctions (Fig.~\ref{fig:slow-modes}d). The effectiveness of the learned representation can also be assessed quantitatively through the implied timescales, which rapidly match the generator reference (Extended Fig.~\ref{fig:extended-1}b).

The capability of recovering the slow modes is not limited to unbiased simulations. In the Supplementary Information, we report how SelfTICA can learn useful dynamical representations also from biased sampling simulations \cite{invernizzi2020rethinking} (Supplementary Fig.~1). In this setting, we have found that using a nonlinear predictor reached a lower training loss more rapidly than a linear one (Supplementary Fig.~2) and more consistently recovered the reference slow modes (Supplementary Fig.~3), without explicit reweighting of time-lagged correlations.

\begin{figure*}[t]  
\centering 
\includegraphics[width=0.95\linewidth]{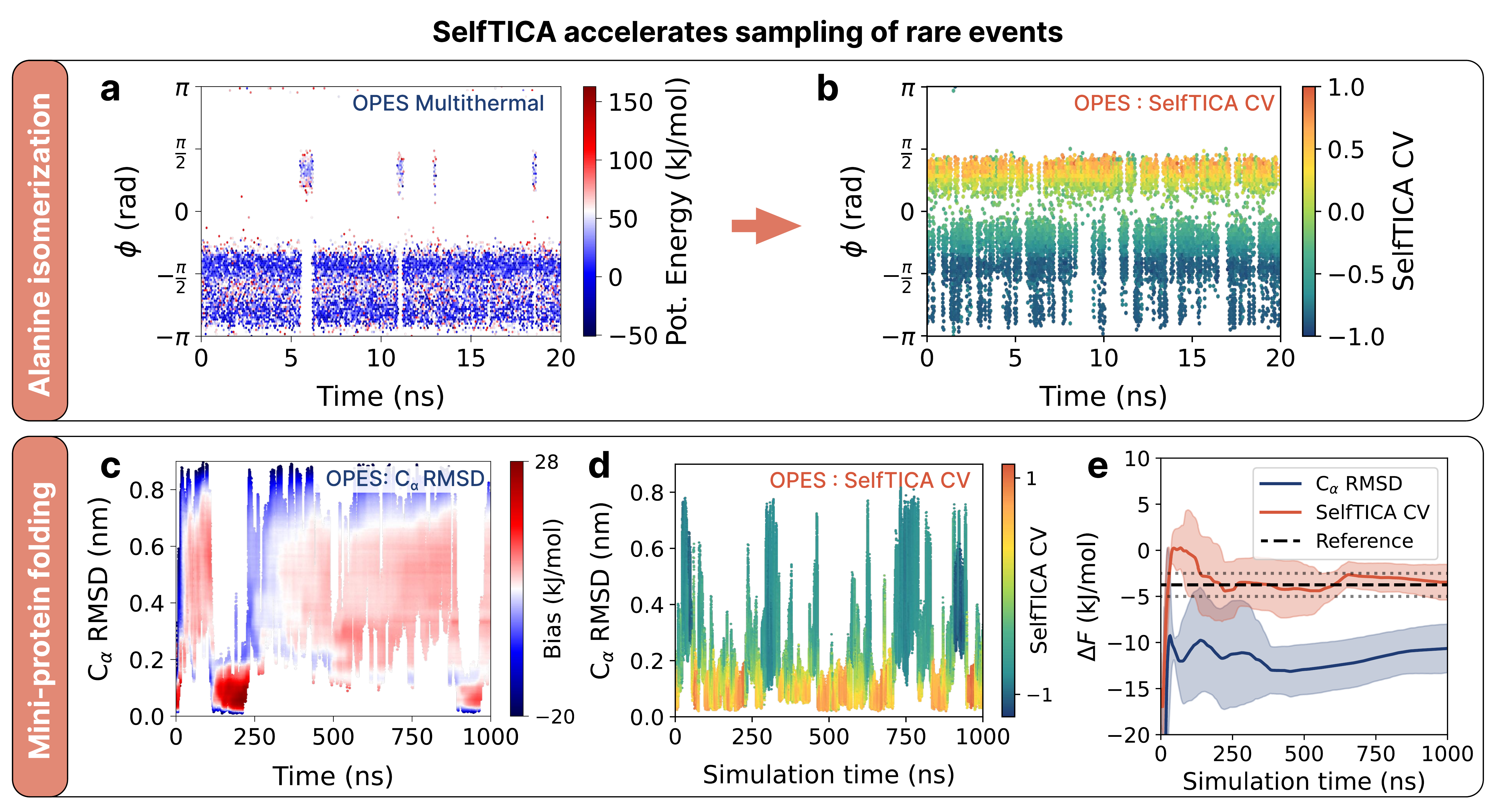} 
\caption{\justifying \textbf{SelfTICA accelerates rare-event sampling.}
(a,b) Alanine dipeptide isomerization. 
(a) Time evolution of the dihedral angle $\phi$ during the OPES-multithermal simulation, showing sparse conformational transitions and colored by the potential energy.
(b) OPES-MetaD simulation biased along the SelfTICA CV learned from the OPES-multithermal trajectory, showing frequent transitions between the relevant conformational regions.
(c--e) Folding of a chignolin mini-protein variant.
(c) OPES-explore trajectory biased by the C$_\alpha$ RMSD, yielding only a few folding--unfolding events.
(d) OPES-explore biased along the learned SelfTICA CV, producing repeated folding and unfolding transitions.
(e) Convergence of the folding free energy difference $\Delta F$, with biasing along the SelfTICA CV approaching the reference value substantially faster and with lower uncertainty than C$_\alpha$ RMSD biasing.
Shaded regions indicate uncertainties from independent simulations.}
\label{fig:Sampling} 
\end{figure*}

\subsection{Improved coordinates over direct optimization methods}

One challenge in constructing CVs from time-lagged trajectories is the choice of lag time, which often requires training and comparing multiple models to identify an appropriate timescale. By separating nonlinear representation learning from spectral decomposition, SelfTICA allows TICA to be applied \textit{a posteriori} at different analysis lag times using the same frozen encoder, at negligible additional computational cost. Extended Fig.~\ref{fig:extended-1}a shows that SelfTICA encoders trained at $\tau_{\mathrm{train}}=1$ or $10$ time units recover the reference eigenvalue spectra across a broad range of analysis lag times. In contrast, an excessively long training lag time of $\tau_{\mathrm{train}}=100$ time units leads to eigenvalue collapse as the relevant dynamical correlations are largely lost.

The same behavior is observed for alanine dipeptide, a well-established benchmark for rare event sampling and free energy calculations. Using an OPES-multithermal trajectory~\cite{invernizzi2020unified} as training data (Supplementary Fig.~6), a SelfTICA encoder trained at the deliberately short lag time $\tau_{\mathrm{train}}=0.1$ ps yields suboptimal CVs when analyzed at the same lag time. However, increasing the TICA analysis lag time to 1 and 10 ps progressively recovers CVs that closely resemble those obtained by directly training SelfTICA at $\tau_{\mathrm{train}}=10$ ps (Extended Fig.~\ref{fig:extended-2}). These results demonstrate that SelfTICA decouples representation learning from the choice of analysis timescale, allowing longer-timescale slow modes to be recovered \textit{a posteriori} without retraining the encoder.

Another significant advantage of SelfTICA with respect to directly optimizing the eigenfunctions and eigenvalues of the dynamical operators is the improved training stability, which results in higher-quality and more robust CVs. For the triple-well potential, across 25 independently trained models, SelfTICA shows substantially lower model-to-model variability than DeepTICA along the minimum free energy path (Supplementary Figs.~4 and 5 and Extended Data Fig.~\ref{fig:extended-3}a), indicating more stable and reproducible slow-mode learning. The same trend is observed for alanine dipeptide across training datasets of increasing size (Supplementary Fig.~7), where SelfTICA consistently exhibits lower CV variability than DeepTICA (Extended Data Fig.~\ref{fig:extended-3}b).

\begin{figure*}[t] 
\centering
\includegraphics[width=1.0\linewidth]{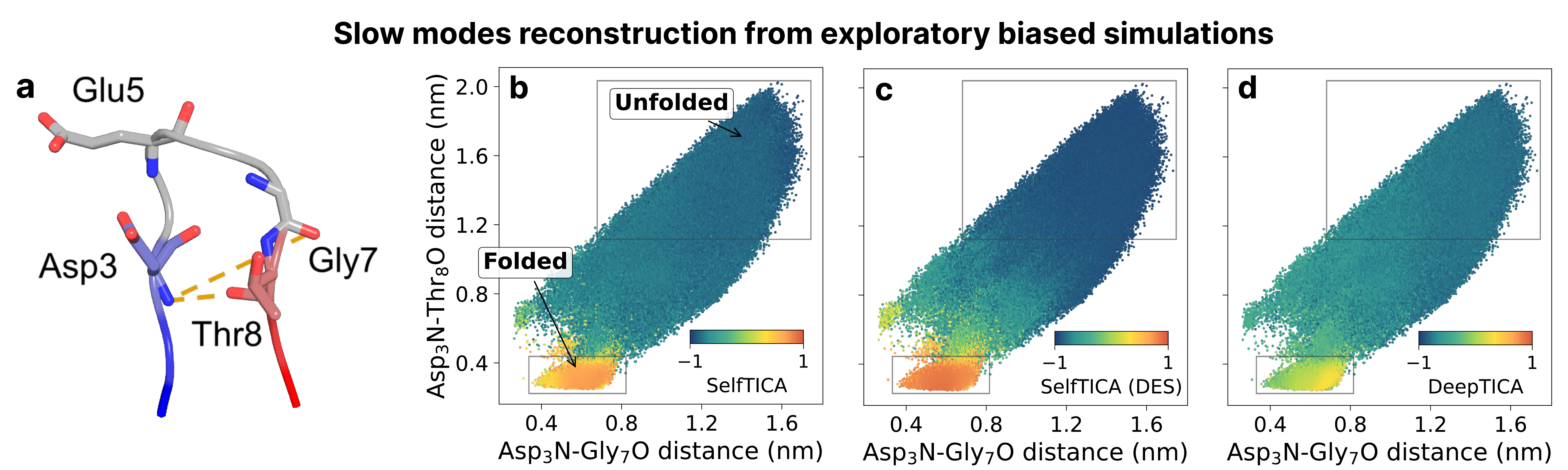}
\caption{\justifying \textbf{SelfTICA reconstructs the reference folding mode of chignolin from exploratory biased simulations.}
(a) Asp3N--Gly7O and Asp3N--Thr8O distances used to project the folding landscape.
(b) SelfTICA trained on the OPES-explore trajectory clearly separates folded and unfolded configurations.
(c) Reference slow mode obtained by training SelfTICA on the long unbiased DESRES trajectory, showing a closely similar organization of the folding landscape.
(d) DeepTICA trained on the same exploratory trajectory provides a less sharply resolved separation. Points are colored by the corresponding learned CV values.}
\label{fig:Chignolin-modes}
\end{figure*}

\subsection{SelfTICA accelerates rare event sampling}

One of the most important requirements of a CV is to be able to accelerate the sampling of the corresponding process. SelfTICA enables learning such coordinates by analyzing exploratory simulations performed without predefined CVs (e.g., via generalized ensemble simulations) or with generic and suboptimal CVs. We illustrate this feature using the isomerization of alanine and the folding of a mini-protein. 

In the first case, we started with an OPES-multithermal simulation~\cite{invernizzi2020unified}, which samples the relevant configurations at a range of temperatures. In such a simulation, transitions between the different conformations remain sparse (Fig.~\ref{fig:Sampling}a). In contrast, biasing the SelfTICA CV trained on such multithermal simulations produces frequent transitions across the relevant regions of the $(\phi,\psi)$ conformational space (Fig.~\ref{fig:Sampling}b). 
Furthermore, the improved robustness and stability of the learned CVs translate into higher-quality biasing coordinates: OPES-MetaD simulations \cite{invernizzi2020rethinking} driven by SelfTICA CVs yield systematically smaller errors in the estimated free energy difference, $\Delta F$, than those driven by similarly trained DeepTICA CVs (Extended Data Fig.~\ref{fig:extended-3}c).

This sampling acceleration also extends to a more complex rare event process such as the folding of (a variant of) the chignolin mini-protein, for which a long unbiased simulation is available ~\cite{lindorff2011fast}. We performed an OPES-explore simulation \cite{invernizzi2022exploration} biased along the C$_\alpha$ RMSD, a notoriously suboptimal CV chosen to generate an aggressively exploratory trajectory. Even with such a biasing scheme, only a couple of folding and unfolding transitions were observed, and the free energy estimate was significantly off from the reference value (Fig.~\ref{fig:Sampling}c). Nevertheless, SelfTICA learns a CV that captures the slow folding dynamics from this exploratory trajectory that can be directly used as a biasing coordinate to accelerate sampling and reconstruct the free energy profile. As illustrated in Fig.~\ref{fig:Sampling}d, OPES-explore simulations driven by the SelfTICA CV exhibit repeated folding and unfolding events, enabling convergence of the folding free energy difference $\Delta F$ in just hundreds of nanoseconds (Fig.~\ref{fig:Sampling}e).

This result is particularly noteworthy because it is achieved using a biased, non-stationary exploratory trajectory generated with a suboptimal CV that only partially captures the underlying slow process. At the same time, this setting is particularly relevant in practice, since a high-quality CV is typically unavailable \textit{a priori} and aggressive sampling is required to promote rare transitions.  Despite being trained only on this imperfect exploratory trajectory, SelfTICA recovers a slow mode that clearly distinguishes the two conformational regions (Fig.~\ref{fig:Chignolin-modes}b), as seen by projecting the sampled configurations onto two structural distances that separate the folded and unfolded states (Fig.~\ref{fig:Chignolin-modes}a). Notably, the resulting slow mode closely resembles that obtained from SelfTICA trained on the long reference unbiased trajectory (Fig.~\ref{fig:Chignolin-modes}c). In contrast, DeepTICA trained on the same exploratory biased trajectory yields a less sharply resolved separation of the slow folding process (Fig.~\ref{fig:Chignolin-modes}d). These results show that SelfTICA can recover the dominant folding mode from challenging exploratory data, providing a practical route to refine initially suboptimal simulations and construct effective coordinates for enhanced sampling.

\begin{figure*}[t] 
\centering
\includegraphics[width=1.0\linewidth]{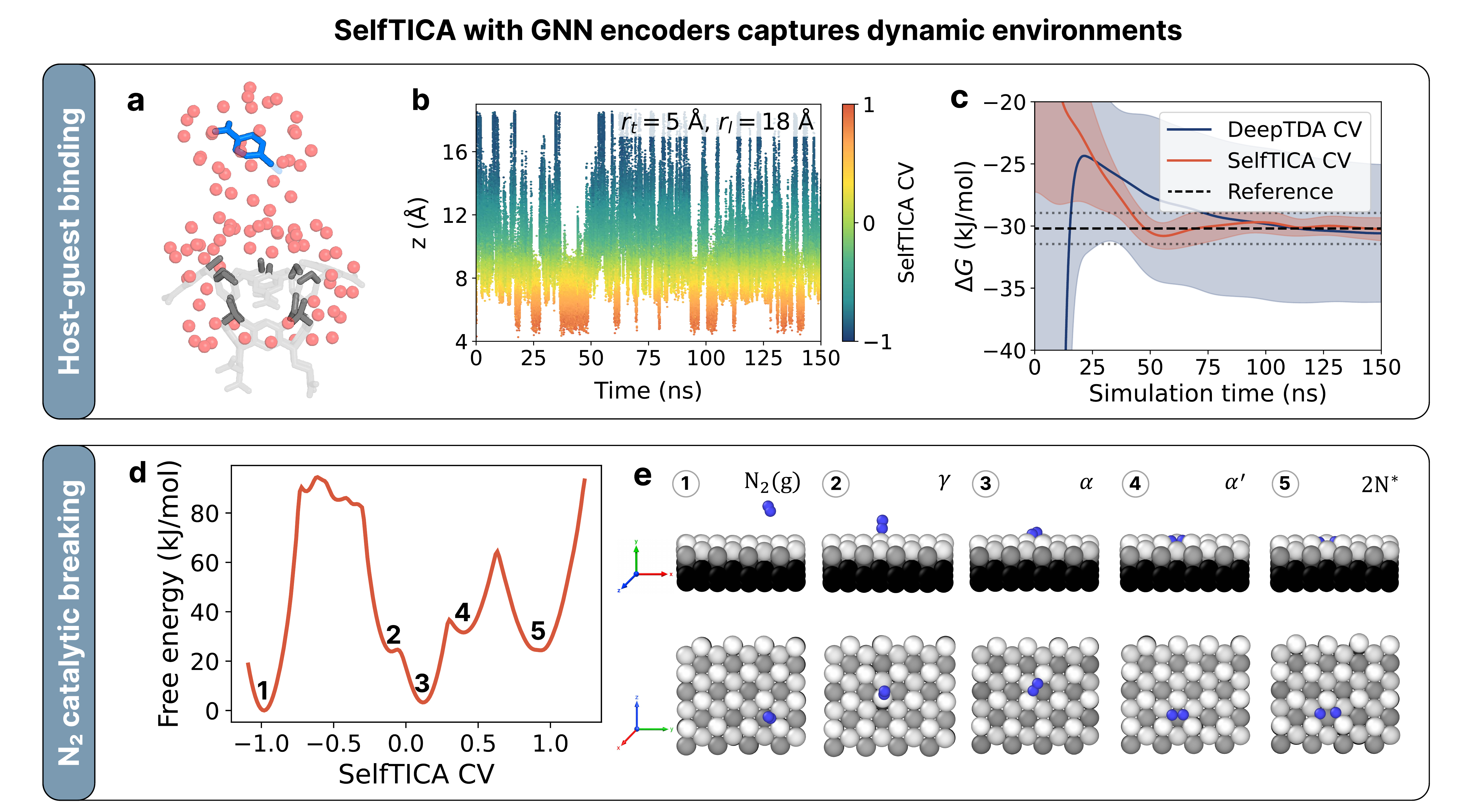}
\caption{\justifying \textbf{GNN-based SelfTICA captures dynamically evolving molecular environments.}
(a--c) OAMe--G2 host--guest binding.
(a) Representative host--guest configuration and surrounding solvent environment used to construct the dynamically updated molecular graph. The truncation radius $r_t$ determines which nearby solvent atoms are included in the graph, while the longer cutoff $r_l$ retains host--guest connections over larger separations.
(b) Time evolution of the host--guest separation $z$ during OPES-MetaD biased along the SelfTICA CV learned from the $r_t=5\,\text{\AA}$ DeepTDA trajectory, with $r_l=18\,\text{\AA}$. Points are colored by the SelfTICA CV and show repeated binding--unbinding transitions.
(c) Convergence of the binding free-energy difference $\Delta G$, with biasing along the SelfTICA CV approaching the reference value faster and with lower uncertainty than DeepTDA biasing.
(d,e) $\mathrm{N_2}$ dissociation on $\mathrm{Fe(111)}$.
(d) Free-energy profile along the learned SelfTICA CV, resolving the major states along the dissociation pathway.
(e) Representative configurations of states 1--5, illustrating adsorption, reorientation, coordination, and final N--N bond cleavage.}
\label{fig:gnn}
\end{figure*}

\subsection{Graph-based encoders capture dynamically changing environments}

Many molecular processes involve local environments whose composition changes as the system evolves, for example through solvent exchange or changes in surface coordination. Fixed handcrafted descriptors are poorly suited to such settings because they generally assume a predefined set of atoms and interactions. To represent such environments, we extended SelfTICA with a GNN encoder that learns directly from atomic structures rather than from a fixed set of handcrafted descriptors. To further increase computational efficiency, we construct the graph by distinguishing between ``system'' atoms, which are always included in the graph, and ``environment'' atoms, which are dynamically added or removed depending on whether they fall within a local cutoff. The resulting graph adapts as solvent molecules or surface atoms enter and leave the interaction region (Extended Data Fig.~\ref{fig:extended-4}).

\begin{figure*}[t] 
\centering
\includegraphics[width=1.0\linewidth]{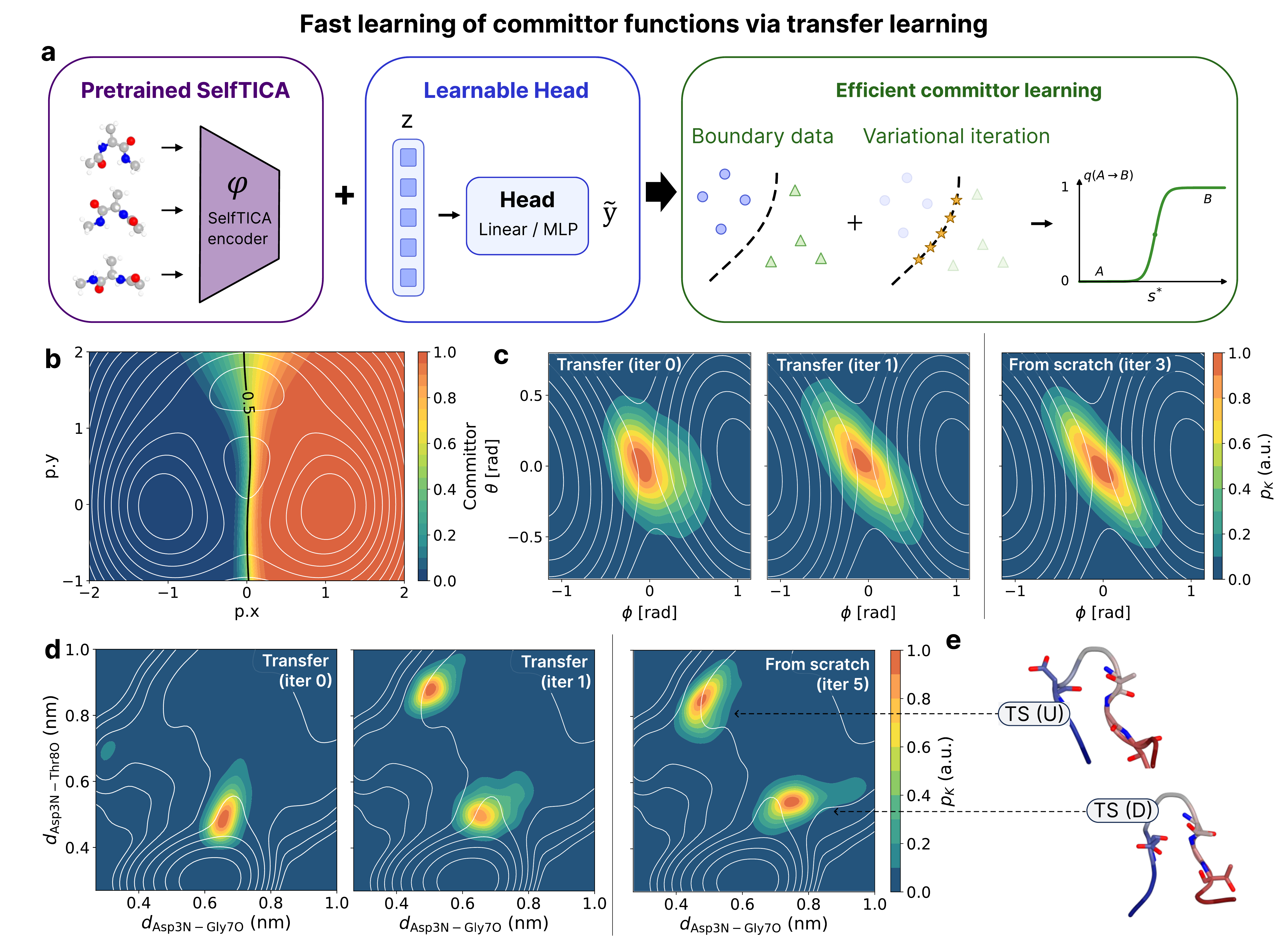}
\caption{\justifying \textbf{Transfer of pretrained SelfTICA representations to committor learning.}
(a) Transfer-learning strategy in which a frozen pretrained SelfTICA encoder is coupled to a lightweight task-specific head for variational committor learning. For alanine dipeptide and chignolin, the head is first trained using boundary-state configurations and then refined once using transition-region configurations generated by committor-guided sampling.
(b) Committor function learned for the triple-well potential using a linear head on the frozen SelfTICA representation.
(c) Kolmogorov distribution, $p_{\mathrm K}$, for alanine dipeptide. After a single transfer refinement, the resulting distribution is comparable to that obtained after three refinement iterations of standard committor training.
(d) $p_{\mathrm K}$ for chignolin. After a single transfer refinement, two transition state regions are resolved that closely resemble those obtained after five refinement iterations of standard committor training.
(e) Representative structures of the two chignolin transition state ensembles, TS(U) and TS(D).}
\label{fig:committor}
\end{figure*}

We first applied the GNN-based encoder to the host--guest binding process OAMe--G2, in which association and dissociation are coupled to rearrangements of the surrounding water network (Fig.~\ref{fig:gnn}a). Explicitly incorporating the solvent environment proved essential: when solvent atoms were excluded, enhanced sampling did not produce efficient reversible binding--unbinding transitions (Supplementary Fig.~11). To determine the spatial extent of solvent information required for an effective CV, we performed OPES-MetaD simulations using a DeepTDA CV while varying the environment cutoff $r_t$ (Extended Data Fig.~\ref{fig:extended-5}). With $r_t=4\,\text{\AA}$, the trajectory remained predominantly confined to the bound region, indicating that the immediate solvation shell alone was insufficient to capture an effective coordinate for host--guest binding. Increasing the cutoff to $r_t=5$ or $6\,\text{\AA}$ enabled repeated transitions between bound and unbound configurations, suggesting that information from solvent molecules beyond the first hydration shell is important for capturing the slow host--guest process.

We therefore trained SelfTICA on the trajectory generated with DeepTDA using $r_t=5\,\text{\AA}$ and subsequently performed OPES-MetaD biasing along the learned CV (Fig.~\ref{fig:gnn}b). This produced frequent binding--unbinding transitions, demonstrating that the dynamical information contained in the DeepTDA trajectory can be distilled into a more effective reaction coordinate. This improvement is also reflected in the binding free energy estimate: simulations driven by the SelfTICA CV rapidly approached the reference $\Delta G$, whereas the classifier-based DeepTDA CV converged more slowly and exhibited substantially larger uncertainty (Fig.~\ref{fig:gnn}c, Supplementary Fig.~12). Thus, the GNN-based SelfTICA representation not only incorporates the dynamically changing solvent environment, but also converts exploratory sampling into a CV that is more effective for thermodynamic sampling.

We next asked whether the same approach could describe the changing coordination environment encountered in heterogeneous catalysis. We considered the dissociation of $\mathrm{N_2}$ on $\mathrm{Fe(111)}$~\cite{bonati2023role}, a multistep reaction involving adsorption, surface-coordination changes, intermediate configurations, and final bond cleavage. This makes it a more stringent test than a simple two-state process for assessing whether the learned representation can resolve complex reaction pathways. SelfTICA was trained on a short 1 ns-long OPES-MetaD simulation biasing geometrical degrees of freedom; see Supplementary Figs.~13--14. The resulting one-dimensional SelfTICA CV organizes the reaction into a sequence of well-resolved metastable and intermediate states from the gas-phase molecular state to the adsorbed intermediates up to the final dissociated configuration (see Fig.~\ref{fig:gnn}d and snapshots in Fig.~\ref{fig:gnn}e). Also in this case, using the learned SelfTICA CV for enhanced sampling simulations results in more transitions between molecular and dissociated configurations than biasing structural coordinates (Supplementary Fig.~15). Despite being learned from short exploratory data, the GNN-based SelfTICA CV therefore also captures coupled structural rearrangements underlying this heterogeneous catalytic reaction.

\subsection{SelfTICA representations accelerate committor learning}

Finally, we show that the representations learned for slow-mode discovery can be reused for a distinct kinetic-learning task. We considered the committor function $q(\mathbf{R})$, defined as the probability that a configuration $\mathbf{R}$ reaches state B before state A. Whereas the slow eigenfunctions learned by SelfTICA describe the global metastable organization of a system, the committor is defined for a selected pair of states and resolves the transition between them, providing a state-specific description of the transition process.

The committor can be learned through a variational principle, but doing so is computationally demanding because it involves higher-order derivatives with respect to molecular coordinates and requires repeated cycles of model training and transition state region sampling~\cite{kang2024computing,trizio2025everything}.  To reduce this cost, we froze the pretrained SelfTICA encoder and optimized only a lightweight task-specific head (Fig.~\ref{fig:committor}a) that maps the learned representation onto a one-dimensional committor space, which is subsequently transformed into the committor probability through a sigmoid function. The head was initialized using configurations from the boundary states (unbiased simulations in states A and B) and subsequently refined using configurations collected through committor-guided sampling~\cite{kang2024computing,trizio2025everything}. If the pretrained representation already contains the relevant kinetic information, this procedure should require fewer refinement cycles than standard committor learning, in which the full model is initialized without pretraining and optimized end-to-end.

We first tested this strategy on the triple-well potential. A linear head was trained on top of the frozen SelfTICA representation using configurations sampled from states A and B to impose the committor boundary conditions, together with configurations from an unbiased trajectory for variational training. The transferred model recovered a smooth committor connecting the two metastable basins, with the $q=0.5$ isocommittor passing through the barrier region (Fig.~\ref{fig:committor}b). Committor-guided sampling using the transferred model produced a Kolmogorov distribution, $p_{\mathrm K}$, strongly enriched around the transition region separating the two basins (Supplementary Fig.~16), confirming that the transferred representation identifies configurations relevant to barrier crossing. Thus, the pretrained SelfTICA representation already contains sufficient kinetic information for a simple linear readout to recover a state-specific reaction coordinate and efficiently guide transition-region sampling.

We next considered alanine dipeptide using the SelfTICA encoder pretrained on the multithermal trajectory. When trained using only boundary-state configurations with a linear head, the transferred model already produced a clearer transition region and a better-defined $q=0.5$ isocommittor than a standard committor model trained on the same boundary-state data without SelfTICA pretraining (Supplementary Fig.~17). Correspondingly, the transferred model was already able to identify transition-relevant configurations, as reflected by the Kolmogorov distribution $p_{\mathrm K}$ (Fig.~\ref{fig:committor}c). After a single sampling and refinement iteration, the committor model became more sharply resolved and $p_{\mathrm K}$ more localized along the transition corridor connecting the metastable basins. The resulting transition state distribution was comparable to that obtained after three refinement iterations when the model was trained from scratch, demonstrating that pretraining substantially reduces the iterative sampling and optimization required for committor learning.

Finally, we applied the same strategy to chignolin folding. Starting from the SelfTICA representation pretrained on the OPES-explore trajectory, we optimized a small nonlinear committor-space head while keeping the encoder fixed; for this system, the nonlinear head achieved a lower variational loss than a linear readout. Even before any refinement, the transferred model provided insight into transition-relevant configurations when projected onto the Asp3N--Gly7O and Asp3N--Thr8O distances (Fig.~\ref{fig:committor}d). After a single refinement iteration, the model resolved two distinct transition state regions closely resembling those obtained after five iterations of training from scratch. This reduced the total sampling required from 5\,$\mu\mathrm{s}$ to 500\,$\mathrm{ns}$. The two regions, denoted TS(U) and TS(D), correspond to structurally distinct transition state ensembles associated with different folding pathways, as illustrated by representative configurations in Fig.~\ref{fig:committor}e.
\section{Discussion}
In this work, we introduced SelfTICA as a novel framework for molecular sampling and free energy calculations. Its central advantage is that, rather than learning a separate nonlinear coordinate for a specific CV objective, SelfTICA uses time-lagged contrastive pretraining to construct a dynamical representation that is subsequently used across different kinetic tasks. This strategy addresses three central challenges in computational molecular science: robust slow modes identification, efficient rare event sampling, and transition-state characterization. First, it separates representation learning from slow mode extraction, improving training robustness while enabling the same frozen representation to recover slow modes across different timescales through \textit{a posteriori} TICA analysis. Second, SelfTICA enables effective CVs to be extracted from challenging exploratory trajectories that are poorly suited to direct slow-mode optimization. The learned modes can thus accelerate the sampling of rare events and enable the estimation of free-energy landscapes. Third, the learned representation provides an excellent basis for the committor function, substantially reducing the iterative sampling and refinement required to locate and characterize the transition-state regions. 

Beyond these methodological advances, the results indicate that SelfTICA is applicable across diverse molecular systems and challenging training data regimes. The learned representations remain useful when training data are limited, biased, or exploratory, and the same learning principle can be extended to dynamically changing molecular environments through graph-based encoders without relying on fixed handcrafted descriptors. Moreover, their successful adaptation to committor learning shows that the encoded dynamical information is not restricted to slow-mode construction or enhanced sampling, but can be used for a distinct kinetic objective. More generally, temporally informative representations may provide a useful basis for studying high-dimensional stochastic systems where labeled data and rare transition events are scarce, potentially extending this paradigm beyond molecular simulation to broader problems in computational science.

Looking ahead, the same representation learning strategy offers a promising route toward transferable models trained on increasingly large and diverse collections of molecular trajectories. Together with expressive architectures capable of spanning different chemical environments ~\cite{ross2022large,fang2022geometry,zhou2023uni}, such pretraining could enable dynamical representations to be adapted to new systems with limited additional sampling, using, for instance, efficient transfer learning techniques~\cite{falk2023transfer,novelli2025fast}. Their utility could also extend beyond the tasks considered here, including kinetic coarse-graining, transfer-operator or generator estimation, rate prediction, and transition-path sampling. By demonstrating that a single dynamics-aware representation can support distinct kinetic objectives, SelfTICA lays the groundwork for general-purpose pretrained models for molecular dynamics.
\section{Methods}
We first describe the construction of time-lagged trajectory datasets used for SelfTICA training, followed by the contrastive representation-learning objective and the encoder architectures used to parameterize the learned feature space. We then discuss the choice of predictor for equilibrium and biased trajectories and establish the connection between the contrastive objective and variational dynamical learning. Finally, we describe post-training TICA analysis for slow-mode extraction, the OPES protocols used for exploratory and enhanced sampling, and the adaptation of frozen SelfTICA representations to committor learning.

\subsection{Trajectory Data Generation for SelfTICA Training}

SelfTICA is trained on time-lagged datasets constructed from molecular dynamics trajectories. Given a trajectory, we sample pairs of configurations separated by a lag time $\tau$, denoted as $(\mathbf{R}_t,\mathbf{R}_{t+\tau})$. Within each mini-batch, the matched pair $(\mathbf{R}_t^{(i)},\mathbf{R}_{t+\tau}^{(i)})$ is treated as a positive pair because it follows the system's $\tau$-time evolution, whereas mismatched pairs $(\mathbf{R}_t^{(i)},\mathbf{R}_{t+\tau}^{(j)})$ with $i\neq j$ are treated as negative pairs because they are not true $\tau$-time successors. This construction enables label-free learning of dynamical representations directly from trajectory data, without requiring predefined metastable states or reaction-coordinate labels. However, the quality and diversity of the learned representation depend strongly on the trajectory used for training.

Ideally, training data are generated from unbiased molecular dynamics simulations, for which the time-lagged pairs directly sample the equilibrium dynamics of the system. For complex molecular systems, however, unbiased simulations may remain trapped in metastable basins and provide insufficient transition information for learning the relevant slow degrees of freedom. We therefore also construct training datasets from trajectories generated using the OPES-based enhanced sampling protocols described below. In this setting, these trajectories are used to provide broad configurational coverage while retaining sufficient information about the relevant slow rearrangements and transition pathways for dynamical representation learning and CV discovery.

\subsection{SelfTICA Representation Learning and Contrastive Objective}

Using the matched and mismatched time-lagged pairs defined above, SelfTICA learns a dynamical representation through a shared encoder $\varphi$ and a predictor $P$. The encoder maps each molecular configuration to a latent representation,
\begin{equation}
\mathbf{z}_t
=
\varphi(\mathbf{R}_t),
\end{equation}
while the predictor approximates its time-lagged evolution,
\begin{equation}
P\mathbf{z}_t
\approx
\mathbf{z}_{t+\tau}.
\end{equation}
The specific encoder architectures and the choice between linear and nonlinear predictors are described in the following subsections.

The similarity between the predicted representation of sample $i$ and the time-lagged representation of sample $j$ is measured through their scalar product,
\begin{equation}
D_{ij}
=
\left\langle
P\mathbf{z}^{(i)}_t,
\mathbf{z}^{(j)}_{t+\tau}
\right\rangle .
\end{equation}
The encoder and predictor are jointly optimized using the contrastive objective~\cite{turri2026selfsupervised,haochen2021provable},
\begin{equation}
\label{eq:em_loss}
\mathcal{L}(\varphi,P)
=
\frac{1}{N(N-1)}
\sum_{i\neq j}D_{ij}^{2}
-
\frac{2}{N}
\sum_{i=1}^{N}D_{ii}.
\end{equation}
The first term suppresses similarity between mismatched configurations, whereas the second promotes agreement between configurations connected by the dynamics. Consequently, optimization encourages the latent representation to retain information that is predictive over the lag time $\tau$ while suppressing correlations between dynamically unrelated configurations.

\subsection{Encoder Architectures}
SelfTICA can be combined with different encoder architectures. When informative descriptors are available, such as pairwise distances or dihedral angles, we use a feed-forward neural network encoder to map them to a multidimensional latent representation.

For systems where handcrafted descriptors are difficult to define, we employ a SchNet-based GNN~\cite{schutt2018schnet,zhang2024descriptor}. A molecular configuration is represented as a graph, where atoms are nodes and interatomic distances are edge features. Edges are constructed within a cutoff radius, defining the neighborhood $\mathcal{N}(i)$ of atom $i$. The initial node features are obtained from atomic numbers and updated through message passing as hidden feature vectors $\mathbf{h}_i^{(l)}$, while geometric information enters through the interatomic distance $r_{ij}$.

At each message-passing layer, neighboring information is aggregated as
\begin{equation}
    \mathcal{M}_i^{(l)}
    =
    \bigoplus_{j\in\mathcal{N}(i)}
    M_{\theta}^{(l)}
    \left(
        \mathbf{h}_i^{(l)},\mathbf{h}_j^{(l)},r_{ij}
    \right),
\end{equation}
where $M_{\theta}^{(l)}$ is a learnable message function and $\bigoplus$ denotes a permutation-invariant aggregation operation. The node feature is then updated by
\begin{equation}
    \mathbf{h}_i^{(l+1)}
    =
    U_{\theta}^{(l)}
    \left(
        \mathbf{h}_i^{(l)},\mathcal{M}_i^{(l)}
    \right).
\end{equation}
where $U_{\theta}^{(l)}$ is a learnable update function that combines the current node feature with the aggregated message. After $L$ message-passing layers, node features are pooled into a graph-level representation,
\begin{equation}
    \mathbf{z}
    =
    R_{\theta}
    \left(
        \rho\left(\{\mathbf{h}_i^{(L)}\}_{i=1}^{N_p}\right)
    \right),
\end{equation}
where $\rho$ is a permutation-invariant pooling function, $N_p$ is the number of pooled nodes, and $R_{\theta}$ is a learnable readout network. The resulting feature vector $\mathbf{z}$ is used for contrastive training and subsequent TICA analysis.

For large systems with complex environments, we use a dual-cutoff truncated graph~\cite{kang2026committors}. The selected atoms are partitioned into system nodes, which define the molecular region of interest, and environment nodes, such as solvent or surface atoms, which provide dynamically changing contextual information. System nodes are always retained, whereas environment nodes are included only when they lie within an effective truncation radius $r_t=r_c+\Delta_b$ of the system nodes, where $r_c$ is the local interaction cutoff and $\Delta_b$ is a buffer introduced for stable neighbor-list updates. Edges involving environment nodes are constructed using the cutoff $r_c$, whereas system--system edges use a larger long-range cutoff $r_l$ to preserve interactions among persistent system atoms. Pooling is restricted to the system nodes, allowing environmental information to influence the learned representation while reducing computational cost and noise arising from intermittently included environment atoms.

\subsection{Variational Interpretation and Predictor Models}

The choice of predictor depends on the type of trajectory used for SelfTICA training. For equilibrium trajectories, as well as quasi-stationary biased ones, we use a linear predictor. Indeed, for equilibrium dynamics, the linear predictor can be interpreted as an approximation to the projected transfer operator in the learned feature space. As shown by Turri \textit{et al.}~\cite{turri2026selfsupervised}, for a fixed encoder $\varphi$ and linear predictor $P$, minimizing the contrastive objective in Eq.~\ref{eq:em_loss} with respect to $P$ gives
\begin{equation}
    P^*
    =
    \arg\min_P \mathcal{L}(\varphi,P)
    =
    C_{00}^{-1} C_{0\tau} C_{\tau\tau}^{-1},
\end{equation}
where $C_{00}=\langle \mathbf{z}_t\mathbf{z}_t^\top\rangle$, $C_{\tau\tau}=\langle \mathbf{z}_{t+\tau}\mathbf{z}_{t+\tau}^\top\rangle$, and $C_{0\tau}=\langle \mathbf{z}_t\mathbf{z}_{t+\tau}^\top\rangle$. Substituting the optimal predictor $P^*$ back into the contrastive objective yields
\begin{equation}
\label{eq:selftica_vamp2}
    \mathcal{L}(\varphi,P^*)
    =
    -
    \left\|
    C_{00}^{-1/2}
    C_{0\tau}
    C_{\tau\tau}^{-1/2}
    \right\|_F^2,
\end{equation}
up to constants independent of the encoder. Thus, minimizing the contrastive objective is equivalent to maximizing the VAMP-2 score~\cite{mardt2018vampnets} for the optimal linear predictor. Furthermore, if the dynamics satisfy detailed balance, stationarity gives $C_{00}=C_{\tau\tau}=C_0$, while reversibility allows $C_{0\tau}=C_{\tau}$. Under these conditions, the VAMP-2 score reduces to the VAC objective,
\begin{equation}
    \mathcal{L}(\varphi,P^*)=-\sum_i\lambda_i^2
\end{equation}
up to encoder-independent constants. Minimizing the contrastive loss therefore encourages the learned feature space to approximate the dominant eigenspace of the reversible transfer operator without directly optimizing individual eigenvalues or eigenfunctions. 

On the other hand, for enhanced sampling-biased trajectories, the applied potentials modify the effective dynamics observed during training. As discussed by Devergne \textit{et al.}~\cite{devergne2024biased,devergne2025slow}, this can be expressed as a perturbation of the infinitesimal generator,
    \begin{equation}
\mathcal{T}_{\tau}=e^{\tau\mathcal{A}},
\qquad  
\mathcal{T}'_{\tau}=e^{\tau\mathcal{A}'},
\qquad
\mathcal{A}'=\mathcal{A}+\Delta\mathcal{A}_{V},
\end{equation}
where $\mathcal{T}_{\tau}$ and $\mathcal{T}'_{\tau}$ are the unbiased and biased transfer operators, $\mathcal{A}$ and $\mathcal{A}'$ are their infinitesimal generators, and $\Delta\mathcal{A}_{V}$ denotes the perturbation induced by the bias potential. Although the bias perturbs the generator additively, its effect on the finite-time transfer operator is nonlinear through the exponential operator. For this reason, we use a small nonlinear neural-network predictor to capture the more complex time-lagged relationships induced by the biased dynamics, especially when the trajectories are obtained with adaptive-bias methods that prioritize exploration over bias convergence, such as OPES-explore (see below), allowing SelfTICA to capture slowly evolving dynamical modes even from the nonstationary trajectories.


\subsection{TICA in the Learned Feature Space for Slow-Mode Analysis and CV Construction}
After optimization, the encoder provides a latent representation $\mathbf{z}=\varphi(\mathbf{R})$ that retains time-lagged dynamical information, but the individual feature coordinates are not necessarily orthogonal slow modes. We therefore apply TICA in the learned feature space to extract slow dynamical modes as linear combinations of the encoder features,
\begin{equation}
    \tilde{\psi}_i(\mathbf{R})
    =
    \mathbf{w}_i^\top \varphi(\mathbf{R}),
\end{equation}
where $\tilde{\psi}_i(\mathbf{R})$ denotes the $i$-th TICA component and $\mathbf{w}_i$ is the corresponding projection vector. Using the same covariance matrices $C_0$ and $C_\tau$, the projection vectors are obtained by solving the generalized eigenvalue problem
\begin{equation}
\label{eq:tica}
    C_{\tau}\mathbf{w}_i
    =
    \tilde{\lambda}_i C_0\mathbf{w}_i.
\end{equation}
This variational problem identifies linear combinations of encoder features with maximal time-lagged autocorrelation. The eigenvectors $\mathbf{w}_i$ define mutually orthogonal projections under the covariance metric. 

The leading TICA components are then used as CVs for enhanced sampling, which in the one-dimensional case corresponds to $s(\mathbf{R})=\tilde{\psi}_1(\mathbf{R})$. For equilibrium or stationary trajectories, the corresponding eigenvalues quantify relaxation times through the implied timescales, $t_i=-\tau/\ln|\tilde{\lambda}_i|$. Because TICA is applied only after contrastive training, its lag time can be varied independently of the contrastive-training lag time, enabling spectra, slow modes, and CVs to be analyzed without retraining the encoder. For biased or non-stationary exploratory trajectories, the TICA components are used primarily as low-dimensional dynamical projections and biasing coordinates, and the associated eigenvalues are not interpreted as unbiased physical relaxation times.

\subsection{Enhanced Sampling Protocols Based on OPES}

Biased or exploratory training trajectories, as well as subsequent enhanced sampling using the learned SelfTICA CVs, are generated using the OPES family of methods. OPES accelerates configurational exploration by constructing a bias potential $V(s)$ that drives the sampled probability distribution toward a prescribed target distribution,
\begin{equation}
V(s) =
\frac{1}{\beta}
\log
\frac{p(s)}{p^{\mathrm{tg}}(s)},
\end{equation}
where $s$ denotes the CVs, $p(s)$ is the probability distribution estimated according to the specific OPES formulation, and $p^{\mathrm{tg}}(s)$ is the target distribution.

Different choices of $p^{\mathrm{tg}}(s)$ define different OPES variants with distinct sampling objectives. OPES-MetaD~\cite{invernizzi2020rethinking} uses a well-tempered~\cite{barducci2008well} target distribution and estimates the unbiased probability through on-the-fly reweighting, thereby progressively building a quasi-static bias that facilitates free-energy estimation along predefined CVs. In contrast, OPES-explore~\cite{invernizzi2022exploration} estimates $p(s)$ from the sampled distribution rather than from the reweighted one, producing a more exploratory bias that helps escape metastable basins and generate diverse transition data when the initial CVs are suboptimal. OPES-multithermal~\cite{invernizzi2020unified} constructs the target distribution as a mixture of overlapping probability distributions associated with different effective temperatures, allowing the simulation to sample configurations relevant over a broad temperature range.

The OPES protocols therefore play two complementary roles in the workflow. First, different OPES simulations provide diverse trajectories from which time-lagged SelfTICA training pairs can be constructed. Second, after the SelfTICA encoder has been trained and slow modes extracted through TICA, the resulting CVs can be used directly in subsequent OPES simulations to enhance the rare events of interest and estimate thermodynamic quantities.

\subsection{Adaptation of Pretrained SelfTICA Representations to Committor Learning}

Given two metastable states $A$ and $B$, the committor $q(\mathbf{R})$ gives the probability that a trajectory initiated from configuration $\mathbf{R}$ reaches $B$ before $A$. The committor can be obtained variationally by minimizing the Kolmogorov functional
\begin{equation}
\mathcal{K}[q]
=
\left\langle
\left|\nabla_{\mathbf{u}} q(\mathbf{R})\right|^2
\right\rangle_U ,
\label{eq:kolmogorov}
\end{equation}
subject to $q(\mathbf{R})=0$ for $\mathbf{R}\in A$ and $q(\mathbf{R})=1$ for $\mathbf{R}\in B$, where $\mathbf{u}$ denotes the mass-weighted Cartesian coordinates and $\langle\cdot\rangle_U$ denotes an equilibrium average over the Boltzmann distribution. Under overdamped dynamics, the minimum of this functional is proportional to the transition rate~\cite{kang2024computing}.

To reuse the pretrained SelfTICA representation for committor learning, we parameterize the committor using a linear readout followed by a sigmoid activation,
\begin{equation}
q_{\theta}(\mathbf{R})
=
\sigma\!\left(
\mathbf{w}^{\top}\varphi(\mathbf{R})+b
\right),
\qquad
\sigma(x)=\frac{1}{1+e^{-x}},
\end{equation}
where \(\varphi\) is the pretrained SelfTICA encoder, \(\theta=\{\mathbf{w},b\}\) denotes the parameters of the task-specific linear readout, and the sigmoid activation constrains the predicted committor to \(0<q_{\theta}(\mathbf{R})<1\). The encoder is kept fixed and only the readout parameters are optimized, allowing the dynamical information learned during SelfTICA pretraining to be directly reused for committor estimation.

The committor model is optimized by minimizing the variational loss
\begin{equation}
\label{eq:committor_var}
\mathcal{L}_{\mathrm{v}}
=
\frac{1}{N}
\sum_{i=1}^{N}
w_i
\left|
\nabla_{\mathbf{u}}
q_{\theta}(\mathbf{R}_i)
\right|^2 ,
\end{equation}
together with the boundary loss
\begin{equation}
\label{eq:committor_boundary}
\mathcal{L}_{\mathrm{b}}
=
\frac{1}{N_A}
\sum_{i\in A}
q_{\theta}(\mathbf{R}_i)^2
+
\frac{1}{N_B}
\sum_{i\in B}
\left(
q_{\theta}(\mathbf{R}_i)-1
\right)^2 .
\end{equation}
The total loss is $\mathcal{L}=\mathcal{L}_{\mathrm{v}}+\alpha\mathcal{L}_{\mathrm{b}}$, where $w_i$ accounts for the statistical weight of configuration $\mathbf{R}_i$ under the sampling protocol and $\alpha$ controls the strength of the boundary constraint.

To enrich configurations in the transition region, we use a regularized and tempered form of the Kolmogorov bias,
\begin{equation}
V_\mathrm{K}(\mathbf{R})
=
-\frac{\lambda}{\beta}
\log
\left[
\left|
\nabla_{\mathbf{u}}
q_{\theta}(\mathbf{R})
\right|^2
+
\epsilon
\right],
\end{equation}
where $\epsilon$ prevents numerical divergence when the committor gradient approaches zero and $\lambda$ controls the bias strength. For $\lambda=1$ and $\epsilon\rightarrow0$, this expression reduces to the original Kolmogorov bias~\cite{kang2024computing}. Starting from a committor head trained on boundary configurations, the current estimate $q_{\theta}$ is used to generate transition-region configurations under this bias. These configurations are incorporated into the subsequent refinement round, while the pretrained SelfTICA encoder remains frozen and only the committor head is updated.

The Kolmogorov distribution corresponding to the unregularized bias is
\begin{equation}
p_{\mathrm{K}}(\mathbf{R})
=
\frac{1}{Z_{\mathrm K}}
e^{-\beta U(\mathbf{R})}
\left|
\nabla_{\mathbf{u}} q(\mathbf{R})
\right|^2,
\label{eq:pk}
\end{equation}
where $Z_{\mathrm K}$ is the normalization constant. This distribution combines the equilibrium probability of a configuration with the local variation of the committor and is therefore concentrated in transition-relevant regions, providing characterization of the transition state ensemble~\cite{kang2024computing}.








\section{DATA AVAILABILITY}
Training and simulation data are available on GitHub \footnotemark[1] and Hugging Face \footnotemark[2].

\section{CODE AVAILABILITY}
The code for training SelfTICA CVs is available through the open-source \texttt{mlcolvar} library \cite{bonati2023unified} alongside didactic tutorials, which is the preferred way to access the most updated code. To obtain the results reported in the manuscript, a frozen version is also available on Hugging Face \footnotemark[2]. The \texttt{PLUMED} \cite{tribello2014plumed,plumed2019promoting} interface for the application of the bias is available on GitHub \footnotemark[1].

\footnotetext[1]{\texttt{GitHub}: \url{https://github.com/Kai-Zhu-2001/SelfTICA/}}
\footnotetext[2]{\textit{Hugging Face}: \url{https://huggingface.co/datasets/Kai-Zhu-2001/SelfTICA}}

\section{Bibliography}
\bibliography{ref}

@book{frenkel2023understanding,
  title={Understanding molecular simulation: from algorithms to applications},
  author={Frenkel, Daan and Smit, Berend},
  year={2023},
  publisher={Elsevier},
  URL={https://www.sciencedirect.com/book/9780323902922/understanding-molecular-simulation}
}

@article{valsson2016enhancing,
  title={Enhancing important fluctuations: Rare events and metadynamics from a conceptual viewpoint},
  author={Valsson, Omar and Tiwary, Pratyush and Parrinello, Michele},
  journal={Annual review of physical chemistry},
  volume={67},
  number={1},
  pages={159--184},
  year={2016},
  publisher={Annual Reviews},
  URL={https://www.annualreviews.org/content/journals/10.1146/annurev-physchem-040215-112229}
}

@article{henin2022Enhanced,
  author    = {Jerome Henin and Tony Lelievre and Michael R. Shirts and Omar Valsson and Lucie Delemotte},
  title     = {Enhanced sampling methods for molecular dynamics simulations},
  journal   = {Living Journal of Computational Molecular Science},
  volume    = {4},
  number    = {1},
  pages     = {1583},
  year      = {2022},
  URL = {https://arxiv.org/abs/2202.04164}
}

@article{barducci2008well,
  title={Well-tempered metadynamics: a smoothly converging and tunable free-energy method},
  author={Barducci, Alessandro and Bussi, Giovanni and Parrinello, Michele},
  journal={Physical review letters},
  volume={100},
  number={2},
  pages={020603},
  year={2008},
  publisher={APS},
  URL={https://journals.aps.org/prl/abstract/10.1103/PhysRevLett.100.020603}
}

@article{valsson2014variational,
  title={Variational approach to enhanced sampling and free energy calculations},
  author={Valsson, Omar and Parrinello, Michele},
  journal={Physical review letters},
  volume={113},
  number={9},
  pages={090601},
  year={2014},
  publisher={APS},
  URL={https://journals.aps.org/prl/abstract/10.1103/PhysRevLett.113.090601}
}

@article{invernizzi2020rethinking,
  title={Rethinking metadynamics: from bias potentials to probability distributions},
  author={Invernizzi, Michele and Parrinello, Michele},
  journal={The journal of physical chemistry letters},
  volume={11},
  number={7},
  pages={2731--2736},
  year={2020},
  publisher={ACS Publications},
  URL={https://pubs.acs.org/doi/abs/10.1021/acs.jpclett.0c00497}
}

@article{invernizzi2020unified,
  title={Unified approach to enhanced sampling},
  author={Invernizzi, Michele and Piaggi, Pablo M and Parrinello, Michele},
  journal={Physical Review X},
  volume={10},
  number={4},
  pages={041034},
  year={2020},
  publisher={APS},
  URL={https://journals.aps.org/prx/abstract/10.1103/PhysRevX.10.041034}
}

@article{invernizzi2022exploration,
  title={Exploration vs convergence speed in adaptive-bias enhanced sampling},
  author={Invernizzi, Michele and Parrinello, Michele},
  journal={Journal of Chemical Theory and Computation},
  volume={18},
  number={6},
  pages={3988--3996},
  year={2022},
  publisher={ACS Publications},
  URL={https://pubs.acs.org/doi/full/10.1021/acs.jctc.2c00152}
}

@article{belkacemi2021chasing,
  title={Chasing collective variables using autoencoders and biased trajectories},
  author={Belkacemi, Zineb and Gkeka, Paraskevi and Leli{\`e}vre, Tony and Stoltz, Gabriel},
  journal={Journal of chemical theory and computation},
  volume={18},
  number={1},
  pages={59--78},
  year={2021},
  publisher={ACS Publications},
  URL={https://pubs.acs.org/doi/abs/10.1021/acs.jctc.1c00415}
}

@article{bonati2020data,
  title={Data-driven collective variables for enhanced sampling},
  author={Bonati, Luigi and Rizzi, Valerio and Parrinello, Michele},
  journal={The journal of physical chemistry letters},
  volume={11},
  number={8},
  pages={2998--3004},
  year={2020},
  publisher={ACS Publications},
  URL={https://pubs.acs.org/doi/abs/10.1021/acs.jpclett.0c00535}
}

@article{trizio2021enhanced,
  title={From enhanced sampling to reaction profiles},
  author={Trizio, Enrico and Parrinello, Michele},
  journal={The Journal of Physical Chemistry Letters},
  volume={12},
  number={35},
  pages={8621--8626},
  year={2021},
  publisher={ACS Publications},
  URL={https://pubs.acs.org/doi/abs/10.1021/acs.jpclett.1c02317}
}

@article{bonati2021deep,
  title={Deep learning the slow modes for rare events sampling},
  author={Bonati, Luigi and Piccini, GiovanniMaria and Parrinello, Michele},
  journal={Proceedings of the National Academy of Sciences},
  volume={118},
  number={44},
  pages={e2113533118},
  year={2021},
  publisher={National Acad Sciences},
  URL={https://www.pnas.org/doi/abs/10.1073/pnas.2113533118}
}

@article{falk2023transfer,
  title={Transfer learning for atomistic simulations using GNNs and kernel mean embeddings},
  author={Falk, John and Bonati, Luigi and Novelli, Pietro and Parrinello, Michele and Pontil, Massimiliano},
  journal={Advances in Neural Information Processing Systems},
  volume={36},
  pages={29783--29797},
  year={2023}
}

@article{kang2024computing,
  title={Computing the committor with the committor to study the transition state ensemble},
  author={Kang, Peilin and Trizio, Enrico and Parrinello, Michele},
  journal={Nature Computational Science},
  pages={1--10},
  year={2024},
  publisher={Nature Publishing Group US New York},
  URL={https://www.nature.com/articles/s43588-024-00645-0}
}

@article{trizio2025everything,
  title={Everything everywhere all at once: a probability-based enhanced sampling approach to rare events},
  author={Trizio, Enrico and Kang, Peilin and Parrinello, Michele},
  journal={Nature Computational Science},
  pages={1--10},
  year={2025},
  publisher={Nature Publishing Group US New York},
  URL={https://www.nature.com/articles/s43588-025-00799-5}
}

@article{chen2019nonlinear,
  title={Nonlinear discovery of slow molecular modes using state-free reversible VAMPnets},
  author={Chen, Wei and Sidky, Hythem and Ferguson, Andrew L},
  journal={The Journal of chemical physics},
  volume={150},
  number={21},
  year={2019},
  publisher={AIP Publishing},
  URL={https://pubs.aip.org/aip/jcp/article/150/21/214114/197931}
}

@article{jung2023machine,
  title={Machine-guided path sampling to discover mechanisms of molecular self-organization},
  author={Jung, Hendrik and Covino, Roberto and Arjun, A and Leitold, Christian and Dellago, Christoph and Bolhuis, Peter G and Hummer, Gerhard},
  journal={Nature Computational Science},
  volume={3},
  number={4},
  pages={334--345},
  year={2023},
  publisher={Nature Publishing Group US New York},
  URL={https://www.nature.com/articles/s43588-023-00428-z}
}

@article{schutt2018schnet,
  title={Schnet--a deep learning architecture for molecules and materials},
  author={Sch{\"u}tt, Kristof T and Sauceda, Huziel E and Kindermans, P-J and Tkatchenko, Alexandre and M{\"u}ller, K-R},
  journal={The Journal of Chemical Physics},
  volume={148},
  number={24},
  year={2018},
  publisher={AIP Publishing},
  URL={https://pubs.aip.org/aip/jcp/article/148/24/241722/962591}
}

@article{shmilovich2023girsanov,
  title={Girsanov Reweighting Enhanced Sampling Technique (GREST): On-the-fly data-driven discovery of and enhanced sampling in slow collective variables},
  author={Shmilovich, Kirill and Ferguson, Andrew L},
  journal={The Journal of Physical Chemistry A},
  volume={127},
  number={15},
  pages={3497--3517},
  year={2023},
  publisher={ACS Publications},
  URL={https://pubs.acs.org/doi/abs/10.1021/acs.jpca.3c00505}
}

@article{prinz2011markov,
  title={Markov models of molecular kinetics: Generation and validation},
  author={Prinz, Jan-Hendrik and Wu, Hao and Sarich, Marco and Keller, Bettina and Senne, Martin and Held, Martin and Chodera, John D and Sch{\"u}tte, Christof and No{\'e}, Frank},
  journal={The Journal of chemical physics},
  volume={134},
  number={17},
  year={2011},
  publisher={AIP Publishing},
  URL={https://pubs.aip.org/aip/jcp/article/134/17/174105/699460}
}

@article{perez2013identification,
  title={Identification of slow molecular order parameters for Markov model construction},
  author={P{\'e}rez-Hern{\'a}ndez, Guillermo and Paul, Fabian and Giorgino, Toni and De Fabritiis, Gianni and No{\'e}, Frank},
  journal={The Journal of chemical physics},
  volume={139},
  number={1},
  year={2013},
  publisher={AIP Publishing},
  URL={https://pubs.aip.org/aip/jcp/article/139/1/015102/192538}
}

@article{nuske2014variational,
  title={Variational approach to molecular kinetics},
  author={Nuske, Feliks and Keller, Bettina G and P{\'e}rez-Hern{\'a}ndez, Guillermo and Mey, Antonia SJS and No{\'e}, Frank},
  journal={Journal of chemical theory and computation},
  volume={10},
  number={4},
  pages={1739--1752},
  year={2014},
  publisher={ACS Publications},
  URL={https://pubs.acs.org/doi/abs/10.1021/ct4009156}
}

@inproceedings{
turri2026selfsupervised,
title={Self-Supervised Evolution Operator Learning for High-Dimensional Dynamical Systems},
author={Giacomo Turri and Luigi Bonati and Kai Zhu and Massimiliano Pontil and Pietro Novelli},
booktitle={The Fourteenth International Conference on Learning Representations},
year={2026},
url={https://openreview.net/forum?id=Ku3kLJle7Q}
}

@article{molgedey1994separation,
  title={Separation of a mixture of independent signals using time delayed correlations},
  author={Molgedey, Lutz and Schuster, Heinz Georg},
  journal={Physical review letters},
  volume={72},
  number={23},
  pages={3634},
  year={1994},
  publisher={APS},
  URL={https://journals.aps.org/prl/abstract/10.1103/PhysRevLett.72.3634}
}

@article{zhang2024descriptor,
  title={Descriptor-Free Collective Variables from Geometric Graph Neural Networks},
  author={Zhang, Jintu and Bonati, Luigi and Trizio, Enrico and Zhang, Odin and Kang, Yu and Hou, TingJun and Parrinello, Michele},
  journal={Journal of Chemical Theory and Computation},
  year={2024},
  publisher={ACS Publications},
  URL={https://pubs.acs.org/doi/abs/10.1021/acs.jctc.4c01197}
}

@article{mardt2018vampnets,
  title={VAMPnets for deep learning of molecular kinetics},
  author={Mardt, Andreas and Pasquali, Luca and Wu, Hao and No{\'e}, Frank},
  journal={Nature communications},
  volume={9},
  number={1},
  pages={5},
  year={2018},
  publisher={Nature Publishing Group UK London},
  URL={https://www.nature.com/articles/s41467-017-02388-1}
}

@article{yang2018refining,
  title={Refining collective coordinates and improving free energy representation in variational enhanced sampling},
  author={Yang, Yi Isaac and Parrinello, Michele},
  journal={Journal of chemical theory and computation},
  volume={14},
  number={6},
  pages={2889--2894},
  year={2018},
  publisher={ACS Publications},
  URL={https://pubs.acs.org/doi/abs/10.1021/acs.jctc.8b00231}
}

@article{lindorff2011fast,
  title={How fast-folding proteins fold},
  author={Lindorff-Larsen, Kresten and Piana, Stefano and Dror, Ron O and Shaw, David E},
  journal={Science},
  volume={334},
  number={6055},
  pages={517--520},
  year={2011},
  publisher={American Association for the Advancement of Science},
  URL={https://www.science.org/doi/10.1126/science.1208351}
}

@article{rizzi2021role,
  title={The role of water in host-guest interaction},
  author={Rizzi, Valerio and Bonati, Luigi and Ansari, Narjes and Parrinello, Michele},
  journal={Nature Communications},
  volume={12},
  number={1},
  pages={93},
  year={2021},
  publisher={Nature Publishing Group UK London},
  URL={https://www.nature.com/articles/s41467-020-20310-0}
}

@article{abraham2015gromacs,
  title={GROMACS: High performance molecular simulations through multi-level parallelism from laptops to supercomputers},
  author={Abraham, Mark James and Murtola, Teemu and Schulz, Roland and P{\'a}ll, Szil{\'a}rd and Smith, Jeremy C and Hess, Berk and Lindahl, Erik},
  journal={SoftwareX},
  volume={1},
  pages={19--25},
  year={2015},
  publisher={Elsevier},
  URL={https://www.sciencedirect.com/science/article/pii/S2352711015000059}
}

@article{tribello2014plumed,
  title={PLUMED 2: New feathers for an old bird},
  author={Tribello, Gareth A and Bonomi, Massimiliano and Branduardi, Davide and Camilloni, Carlo and Bussi, Giovanni},
  journal={Computer physics communications},
  volume={185},
  number={2},
  pages={604--613},
  year={2014},
  publisher={Elsevier},
  URL={https://www.sciencedirect.com/science/article/pii/S0010465513003196}
}

@article{plumed2019promoting,
  title={Promoting transparency and reproducibility in enhanced molecular simulations},
  journal={Nature methods},
  volume={16},
  number={8},
  pages={670--673},
  year={2019},
  publisher={Nature Publishing Group US New York},
  URL={https://www.nature.com/articles/s41592-019-0506-8}
}

@article{salomon2013overview,
  title={An overview of the Amber biomolecular simulation package},
  author={Salomon-Ferrer, Romelia and Case, David A and Walker, Ross C},
  journal={Wiley Interdisciplinary Reviews: Computational Molecular Science},
  volume={3},
  number={2},
  pages={198--210},
  year={2013},
  publisher={Wiley Online Library},
  URL={https://wires.onlinelibrary.wiley.com/doi/abs/10.1002/wcms.1121}
}

@article{piana2011robust,
  title={How robust are protein folding simulations with respect to force field parameterization?},
  author={Piana, Stefano and Lindorff-Larsen, Kresten and Shaw, David E},
  journal={Biophysical journal},
  volume={100},
  number={9},
  pages={L47--L49},
  year={2011},
  publisher={Elsevier},
  URL={https://www.cell.com/biophysj/fulltext/S0006-3495(11)00409-7}
}

@article{mackerell1998all,
  title={All-atom empirical potential for molecular modeling and dynamics studies of proteins},
  author={MacKerell Jr, Alex D and Bashford, Donald and Bellott, MLDR and Dunbrack Jr, Roland Leslie and Evanseck, Jeffrey D and Field, Martin J and Fischer, Stefan and Gao, Jiali and Guo, Houyang and Ha, Sookhee and others},
  journal={The journal of physical chemistry B},
  volume={102},
  number={18},
  pages={3586--3616},
  year={1998},
  publisher={ACS Publications},
  URL={https://pubs.acs.org/doi/abs/10.1021/jp973084f}
}

@article{wang2004development,
  title={Development and testing of a general amber force field},
  author={Wang, Junmei and Wolf, Romain M and Caldwell, James W and Kollman, Peter A and Case, David A},
  journal={Journal of computational chemistry},
  volume={25},
  number={9},
  pages={1157--1174},
  year={2004},
  publisher={Wiley Online Library},
  URL={https://onlinelibrary.wiley.com/doi/abs/10.1002/jcc.20035}
}

@article{bussi2007canonical,
  title={Canonical sampling through velocity rescaling},
  author={Bussi, Giovanni and Donadio, Davide and Parrinello, Michele},
  journal={The Journal of chemical physics},
  volume={126},
  number={1},
  year={2007},
  publisher={AIP Publishing},
  URL={https://pubs.aip.org/aip/jcp/article-abstract/126/1/014101/186581/Canonical-sampling-through-velocity-rescaling?redirectedFrom=fulltext}
}

@article{limongelli2013funnel,
  title={Funnel metadynamics as accurate binding free-energy method},
  author={Limongelli, Vittorio and Bonomi, Massimiliano and Parrinello, Michele},
  journal={Proceedings of the National Academy of Sciences},
  volume={110},
  number={16},
  pages={6358--6363},
  year={2013},
  publisher={National Academy of Sciences},
  URL={https://www.pnas.org/doi/abs/10.1073/pnas.1303186110}
}

@article{bhakat2017resolving,
  title={Resolving the problem of trapped water in binding cavities: prediction of host--guest binding free energies in the SAMPL5 challenge by funnel metadynamics},
  author={Bhakat, Soumendranath and S{\"o}derhjelm, P{\"a}r},
  journal={Journal of computer-aided molecular design},
  volume={31},
  pages={119--132},
  year={2017},
  publisher={Springer},
  URL={https://link.springer.com/article/10.1007/s10822-016-9948-6}
}

@article{bussi2007accurate,
  title={Accurate sampling using Langevin dynamics},
  author={Bussi, Giovanni and Parrinello, Michele},
  journal={Physical Review E—Statistical, Nonlinear, and Soft Matter Physics},
  volume={75},
  number={5},
  pages={056707},
  year={2007},
  publisher={APS},
  URL={https://journals.aps.org/pre/abstract/10.1103/PhysRevE.75.056707}
}

@article{megias2025iterative,
    author={Megías, Alberto and Contreras Arredondo, Sergio and Chen, Cheng Giuseppe and Tang, Chenyu and Roux, Benoît and Chipot, Christophe} ,
    title = {Iterative variational learning of committor-consistent transition pathways using artificial neural networks},
    journal = {Nature Computational Science},
    year = {2025},
    URL={https://www.nature.com/articles/s43588-025-00828-3}
}

@article{wang2021state,
  title={State predictive information bottleneck},
  author={Wang, Dedi and Tiwary, Pratyush},
  journal={The Journal of Chemical Physics},
  volume={154},
  number={13},
  year={2021},
  publisher={AIP Publishing},
  URL={https://pubs.aip.org/aip/jcp/article/154/13/134111/1013207}
}

@article{mehdi2024enhanced,
  title={Enhanced sampling with machine learning},
  author={Mehdi, Shams and Smith, Zachary and Herron, Lukas and Zou, Ziyue and Tiwary, Pratyush},
  journal={Annual Review of Physical Chemistry},
  volume={75},
  number={2024},
  pages={347--370},
  year={2024},
  publisher={Annual Reviews},
  URL={https://doi.org/10.1146/annurev-physchem-083122-125941}
}

@article{zhu2025enhanced,
  title={Enhanced Sampling in the Age of Machine Learning: Algorithms and Applications},
  author={Zhu, Kai and Trizio, Enrico and Zhang, Jintu and Hu, Renling and Jiang, Linlong and Hou, Tingjun and Bonati, Luigi},
  journal={Chemical Reviews},
  year={2025},
  publisher={ACS Publications},
  URL={https://doi.org/10.1021/acs.chemrev.5c00700}
}

@article{novelli2025fast,
  title={Fast and Fourier features for transfer learning of interatomic potentials},
  author={Novelli, Pietro and Meanti, Giacomo and Buigues, Pedro J and Rosasco, Lorenzo and Parrinello, Michele and Pontil, Massimiliano and Bonati, Luigi},
  journal={npj Computational Materials},
  volume={11},
  number={1},
  pages={293},
  year={2025},
  publisher={Nature Publishing Group UK London},
  URL={https://www.nature.com/articles/s41524-025-01779-z}
}

@article{bonati2023unified,
  title={A unified framework for machine learning collective variables for enhanced sampling simulations: mlcolvar},
  author={Bonati, Luigi and Trizio, Enrico and Rizzi, Andrea and Parrinello, Michele},
  journal={The Journal of Chemical Physics},
  volume={159},
  number={1},
  year={2023},
  publisher={AIP Publishing},
  URL={https://pubs.aip.org/aip/jcp/article/159/1/014801/2901354}
}

@article{kang2026committors,
  title={Committors without Descriptors},
  author={Kang, Peilin and Zhang, Jintu and Trizio, Enrico and Hou, TingJun and Parrinello, Michele},
  journal={Journal of Chemical Theory and Computation},
  year={2026},
  publisher={ACS Publications},
  url={https://pubs.acs.org/doi/full/10.1021/acs.jctc.5c01848}
}

@article{darden1993particle,
  title={Particle mesh Ewald: An N  log (N) method for Ewald sums in large systems},
  author={Darden, Tom and York, Darrin and Pedersen, Lee and others},
  journal={Journal of chemical physics},
  volume={98},
  number={12},
  pages={10089--10092},
  year={1993},
  publisher={American Institute of Physics},
  url={http://dx.doi.org/10.1063/1.464397}
}

@article{hess1997lincs,
  title={LINCS: A linear constraint solver for molecular simulations},
  author={Hess, Berk and Bekker, Henk and Berendsen, Herman JC and Fraaije, Johannes GEM},
  journal={Journal of computational chemistry},
  volume={18},
  number={12},
  pages={1463--1472},
  year={1997},
  publisher={Wiley Online Library},
  url={https://doi.org/10.1002/(SICI)1096-987X(199709)18:12%3C1463::AID-JCC4%3E3.0.CO;2-H}
}

@article{bonati2023role,
  title={The role of dynamics in heterogeneous catalysis: Surface diffusivity and N2 decomposition on Fe (111)},
  author={Bonati, Luigi and Polino, Daniela and Pizzolitto, Cristina and Biasi, Pierdomenico and Eckert, Rene and Reitmeier, Stephan and Schl{\"o}gl, Robert and Parrinello, Michele},
  journal={Proceedings of the National Academy of Sciences},
  volume={120},
  number={50},
  pages={e2313023120},
  year={2023},
  publisher={National Academy of Sciences},
  url={https://doi.org/10.1073/pnas.231302312}
}

@article{thompson2022lammps,
  title={LAMMPS-a flexible simulation tool for particle-based materials modeling at the atomic, meso, and continuum scales},
  author={Thompson, Aidan P and Aktulga, H Metin and Berger, Richard and Bolintineanu, Dan S and Brown, W Michael and Crozier, Paul S and In't Veld, Pieter J and Kohlmeyer, Axel and Moore, Stan G and Nguyen, Trung Dac and others},
  journal={Computer physics communications},
  volume={271},
  pages={108171},
  year={2022},
  publisher={Elsevier},
  url={https://doi.org/10.1016/j.cpc.2021.108171}
}

@inproceedings{batatia2022mace,
 author = {Batatia, Ilyes and Kovacs, David P and Simm, Gregor and Ortner, Christoph and Csanyi, Gabor},
 booktitle = {Advances in Neural Information Processing Systems},
 editor = {S. Koyejo and S. Mohamed and A. Agarwal and D. Belgrave and K. Cho and A. Oh},
 pages = {11423--11436},
 publisher = {Curran Associates, Inc.},
 title = {MACE: Higher Order Equivariant Message Passing Neural Networks for Fast and Accurate Force Fields},
 url = {https://proceedings.neurips.cc/paper_files/paper/2022/file/4a36c3c51af11ed9f34615b81edb5bbc-Paper-Conference.pdf},
 volume = {35},
 year = {2022}
}

@article{perego2024data,
  title={Data efficient machine learning potentials for modeling catalytic reactivity via active learning and enhanced sampling},
  author={Perego, Simone and Bonati, Luigi},
  journal={npj Computational Materials},
  volume={10},
  number={1},
  pages={291},
  year={2024},
  publisher={Nature Publishing Group UK London},
  url={https://doi.org/10.1038/s41524-024-01481-6}
}

@inproceedings{devergne2024biased,
  title     = {From Biased to Unbiased Dynamics: An Infinitesimal Generator Approach},
  author    = {Devergne, Timoth{\'e}e and Kostic, Vladimir R. and Parrinello, Michele and Pontil, Massimiliano},
  booktitle = {Advances in Neural Information Processing Systems},
  year      = {2024},
  url={https://doi.org/10.52202/079017-2404}
}

@article{devergne2025slow,
  title={Slow dynamical modes from static averages},
  author={Devergne, Timoth{\'e}e and Kostic, Vladimir and Pontil, Massimiliano and Parrinello, Michele},
  journal={The Journal of Chemical Physics},
  volume={162},
  number={12},
  year={2025},
  publisher={AIP Publishing},
  url={https://doi.org/10.1063/5.0246248}
}

@article{wu2017variational,
  title={Variational Koopman models: Slow collective variables and molecular kinetics from short off-equilibrium simulations},
  author={Wu, Hao and N{\"u}ske, Feliks and Paul, Fabian and Klus, Stefan and Koltai, P{\'e}ter and No{\'e}, Frank},
  journal={The Journal of chemical physics},
  volume={146},
  number={15},
  year={2017},
  publisher={AIP Publishing},
  url={https://doi.org/10.1063/1.4979344}
}

@article{wang2019past,
  title={Past--future information bottleneck for sampling molecular reaction coordinate simultaneously with thermodynamics and kinetics},
  author={Wang, Yihang and Ribeiro, Jo{\~a}o Marcelo Lamim and Tiwary, Pratyush},
  journal={Nature communications},
  volume={10},
  number={1},
  pages={3573},
  year={2019},
  publisher={Nature Publishing Group UK London},
  url={https://www.nature.com/articles/s41467-019-11405-4}
}

@article{chen2018molecular,
  title={Molecular enhanced sampling with autoencoders: On-the-fly collective variable discovery and accelerated free energy landscape exploration},
  author={Chen, Wei and Ferguson, Andrew L},
  journal={Journal of computational chemistry},
  volume={39},
  number={25},
  pages={2079--2102},
  year={2018},
  publisher={Wiley Online Library},
  url={https://doi.org/10.1002/jcc.25520}
}

@article{chen2018collective,
  title={Collective variable discovery and enhanced sampling using autoencoders: Innovations in network architecture and error function design},
  author={Chen, Wei and Tan, Aik Rui and Ferguson, Andrew L},
  journal={The Journal of chemical physics},
  volume={149},
  number={7},
  year={2018},
  publisher={AIP Publishing},
  url={https://doi.org/10.1063/1.5023804}
}

@article{vanden2010transition,
  title={Transition-path theory and path-finding algorithms for the study of rare events.},
  author={Vanden-Eijnden, Eric and others},
  journal={Annual review of physical chemistry},
  volume={61},
  pages={391--420},
  year={2010},
  url={https://doi.org/10.1146/annurev.physchem.040808.090412}
}

@inproceedings{haochen2021provable,
  title     = {Provable Guarantees for Self-Supervised Deep Learning with Spectral Contrastive Loss},
  author    = {HaoChen, Jeff Z. and Wei, Colin and Gaidon, Adrien and Ma, Tengyu},
  booktitle = {Advances in Neural Information Processing Systems},
  volume    = {34},
  pages     = {5000--5011},
  year      = {2021},
  publisher = {Curran Associates, Inc.},
  url={https://proceedings.neurips.cc/paper_files/paper/2021/file/27debb435021eb68b3965290b5e24c49-Paper.pdf}
}

@InProceedings{chen2020simple,
  title = 	 {A Simple Framework for Contrastive Learning of Visual Representations},
  author =       {Chen, Ting and Kornblith, Simon and Norouzi, Mohammad and Hinton, Geoffrey},
  booktitle = 	 {Proceedings of the 37th International Conference on Machine Learning},
  pages = 	 {1597--1607},
  year = 	 {2020},
  editor = 	 {III, Hal Daumé and Singh, Aarti},
  volume = 	 {119},
  series = 	 {Proceedings of Machine Learning Research},
  month = 	 {13--18 Jul},
  publisher =    {PMLR},
  url = 	 {https://proceedings.mlr.press/v119/chen20j.html}
}

@article{wang2022molecular,
  title={Molecular contrastive learning of representations via graph neural networks},
  author={Wang, Yuyang and Wang, Jianren and Cao, Zhonglin and Barati Farimani, Amir},
  journal={Nature Machine Intelligence},
  volume={4},
  number={3},
  pages={279--287},
  year={2022},
  publisher={Nature Publishing Group UK London},
  url={https://www.nature.com/articles/s42256-022-00447-x}
}

@InProceedings{he2020momentum,
author = {He, Kaiming and Fan, Haoqi and Wu, Yuxin and Xie, Saining and Girshick, Ross},
title = {Momentum Contrast for Unsupervised Visual Representation Learning},
booktitle = {Proceedings of the IEEE/CVF Conference on Computer Vision and Pattern Recognition (CVPR)},
month = {June},
year = {2020},
url={https://openaccess.thecvf.com/content_CVPR_2020/papers/He_Momentum_Contrast_for_Unsupervised_Visual_Representation_Learning_CVPR_2020_paper.pdf}
}

@InProceedings{radford2021learning,
  title = 	 {Learning Transferable Visual Models From Natural Language Supervision},
  author =       {Radford, Alec and Kim, Jong Wook and Hallacy, Chris and Ramesh, Aditya and Goh, Gabriel and Agarwal, Sandhini and Sastry, Girish and Askell, Amanda and Mishkin, Pamela and Clark, Jack and Krueger, Gretchen and Sutskever, Ilya},
  booktitle = 	 {Proceedings of the 38th International Conference on Machine Learning},
  pages = 	 {8748--8763},
  year = 	 {2021},
  editor = 	 {Meila, Marina and Zhang, Tong},
  volume = 	 {139},
  series = 	 {Proceedings of Machine Learning Research},
  month = 	 {18--24 Jul},
  publisher =    {PMLR},
  url = 	 {https://proceedings.mlr.press/v139/radford21a.html}
}

@article{zeng2022accurate,
  title={Accurate prediction of molecular properties and drug targets using a self-supervised image representation learning framework},
  author={Zeng, Xiangxiang and Xiang, Hongxin and Yu, Linhui and Wang, Jianmin and Li, Kenli and Nussinov, Ruth and Cheng, Feixiong},
  journal={Nature Machine Intelligence},
  volume={4},
  number={11},
  pages={1004--1016},
  year={2022},
  publisher={Nature Publishing Group UK London},
  url={https://doi.org/10.1038/s42256-022-00557-6}
}

@inproceedings{zhou2023uni,
  title={Uni-mol: A universal 3d molecular representation learning framework},
  author={Zhou, Gengmo and Gao, Zhifeng and Ding, Qiankun and Zheng, Hang and Xu, Hongteng and Wei, Zhewei and Zhang, Linfeng and Ke, Guolin},
  booktitle={The eleventh international conference on learning representations},
  year={2023},
  url={https://openreview.net/forum?id=6K2RM6wVqKu}
}

@article{ross2022large,
  title={Large-scale chemical language representations capture molecular structure and properties},
  author={Ross, Jerret and Belgodere, Brian and Chenthamarakshan, Vijil and Padhi, Inkit and Mroueh, Youssef and Das, Payel},
  journal={Nature Machine Intelligence},
  volume={4},
  number={12},
  pages={1256--1264},
  year={2022},
  publisher={Nature Publishing Group UK London},
  url={https://www.nature.com/articles/s42256-022-00580-7}
}

@article{fang2022geometry,
  title={Geometry-enhanced molecular representation learning for property prediction},
  author={Fang, Xiaomin and Liu, Lihang and Lei, Jieqiong and He, Donglong and Zhang, Shanzhuo and Zhou, Jingbo and Wang, Fan and Wu, Hua and Wang, Haifeng},
  journal={Nature Machine Intelligence},
  volume={4},
  number={2},
  pages={127--134},
  year={2022},
  publisher={Nature Publishing Group UK London}
}

@article{wang2019backpropagation,
  title={Backpropagation-Friendly Eigendecomposition},
  author={Wei Wang and Zheng Dang and Yinlin Hu and Pascal V. Fua and Mathieu Salzmann},
  journal={ArXiv},
  year={2019},
  volume={abs/1906.09023},
  url={https://api.semanticscholar.org/CorpusID:195316577}
}

@ARTICLE{wang2021robust,
  author={Wang, Wei and Dang, Zheng and Hu, Yinlin and Fua, Pascal and Salzmann, Mathieu},
  journal={IEEE Transactions on Pattern Analysis and Machine Intelligence}, 
  title={Robust Differentiable SVD}, 
  year={2022},
  volume={44},
  number={9},
  pages={5472-5487},
  doi={10.1109/TPAMI.2021.3072422}}

\section{ACKNOWLEDGMENTS}
We are grateful to T. Devergne, E. Trizio, G. Rossi, and U. Raucci for helpful discussions and feedback on the manuscript. K.Z. gratefully acknowledges support from Zhejiang University and thanks R. Hu for providing computational resources. We also acknowledge the CINECA award under the ISCRA initiative for providing access to high-performance computing resources and support.

\section{FUNDING}
The authors disclose support for the research of this work from the National Key Research and Development Program of China [grant number 2025ZD1803103].
\section{AUTHOR CONTRIBUTIONS}
K.Z. and L.B. designed the study and T.H. and L.B. supervised the project. K.Z. and P.N. developed the SelfTICA code, while J.Z. and K.Z. developed the GNN-related code. K.Z. performed all model training and simulations. All authors analyzed and discussed the results. K.Z. and J.Z. drafted the initial version of the manuscript, and all authors contributed to the revision and editing.
\section{Competing Interests}
The authors declare no competing interests.

\clearpage
\begin{onecolumngrid}
\captionsetup[figure]{
    format=plain,
    labelfont=bf
}
\renewcommand{\figurename}{Extended Data Fig.}
\setcounter{figure}{0}

\begin{figure}[H] 
\centering
\includegraphics[width=0.75\linewidth]{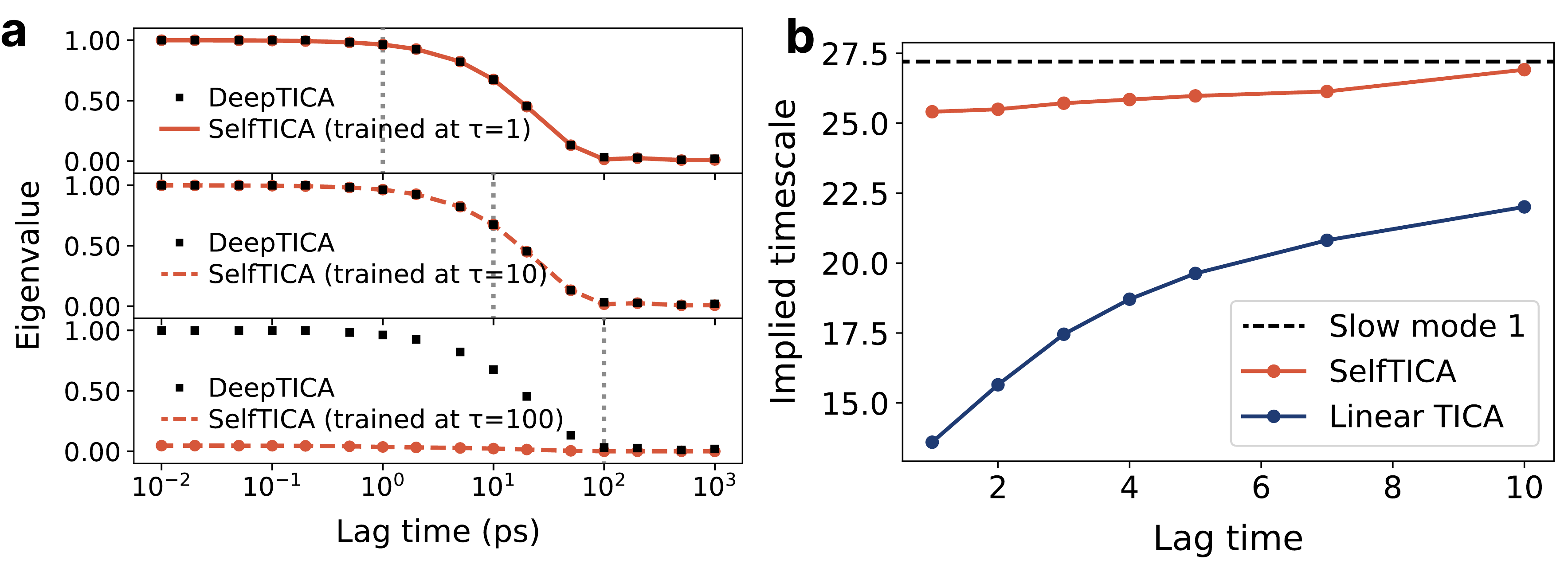}
\caption{\justifying \textbf{Lag-time transfer of SelfTICA representations.}
(a) Leading eigenvalue evaluated over different TICA lag times using SelfTICA encoders trained at $\tau=1$, $10$, and $100$ ps, compared with DeepTICA models trained separately at each different lag time. Vertical dotted lines mark the corresponding training lag times.
(b) Implied timescale of the dominant slow mode obtained with SelfTICA and linear TICA. The dashed line indicates the reference value, which is recovered more rapidly by SelfTICA as the lag time increases.}
\label{fig:extended-1}
\end{figure}

\begin{figure}[H] 
\centering
\includegraphics[width=1.0\linewidth]{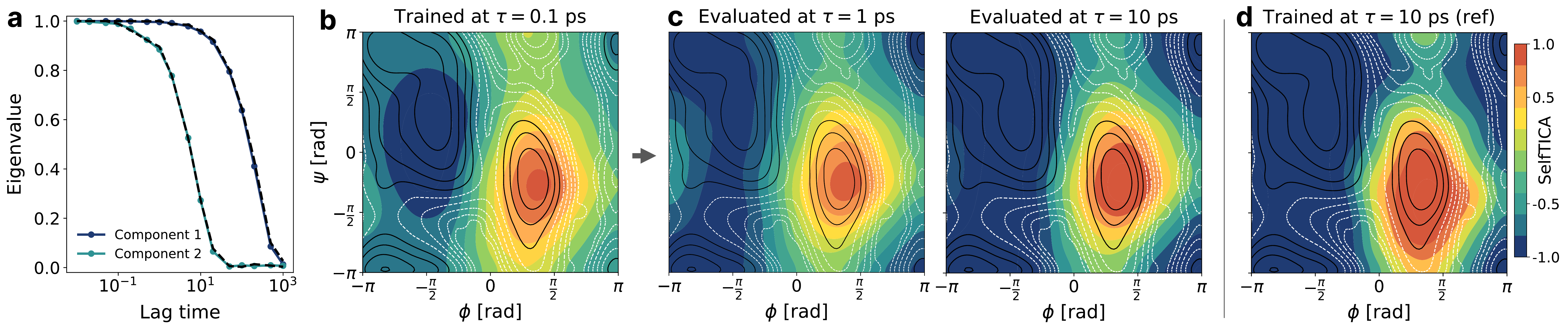}
\caption{\justifying \textbf{Lag-time transfer of SelfTICA for alanine dipeptide.}
(a) Leading TICA eigenvalues evaluated as a function of lag time from a SelfTICA representation learned on the multithermal trajectory.
(b) CV obtained from an encoder trained and evaluated at $\tau=0.1$ ps.
(c) CVs obtained from the same frozen encoder by increasing the TICA lag time to 1 and 10 ps.
(d) Reference CV obtained by directly training SelfTICA at $\tau=10$ ps. Increasing the analysis lag time progressively recovers the long-timescale CV without retraining the encoder.}
\label{fig:extended-2}
\end{figure}

\begin{figure}[H] 
\centering
\includegraphics[width=1.0\linewidth]{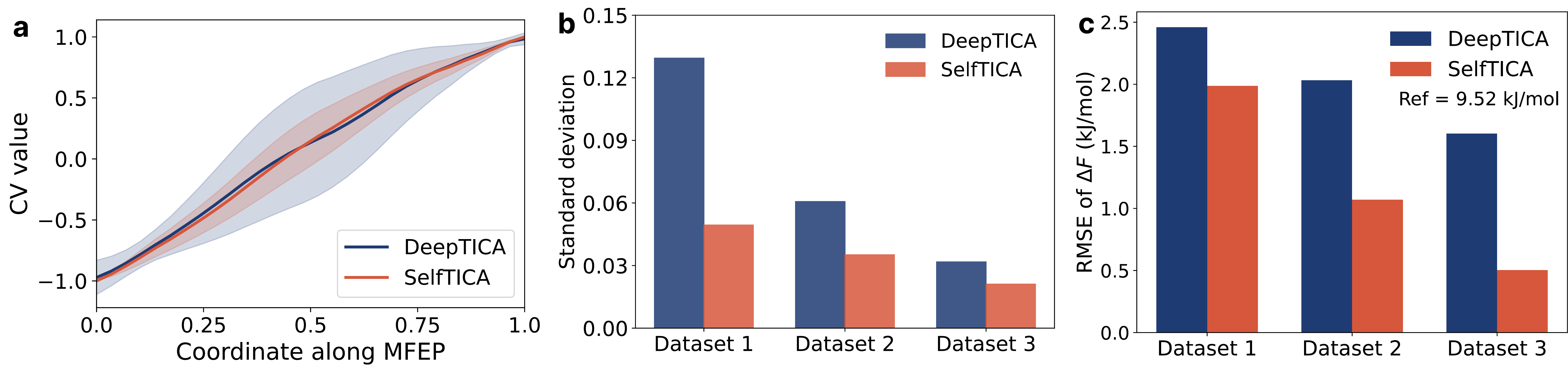}
\caption{\justifying \textbf{Improved robustness of SelfTICA translates into more accurate free-energy estimates.}
(a) For the tri-well potential, CV profiles along the minimum free-energy path obtained with DeepTICA and SelfTICA, with shaded regions indicating variability.
(b,c) For alanine dipeptide, (b) standard deviation of the learned CVs across three datasets and (c) RMSE of the corresponding free-energy difference $\Delta F$ relative to the reference value of $9.52\,\mathrm{kJ\,mol^{-1}}$. SelfTICA exhibits lower variability and consistently smaller free-energy errors than DeepTICA.}
\label{fig:extended-3}
\end{figure}

\begin{figure}[H] 
\centering
\includegraphics[width=0.8\linewidth]{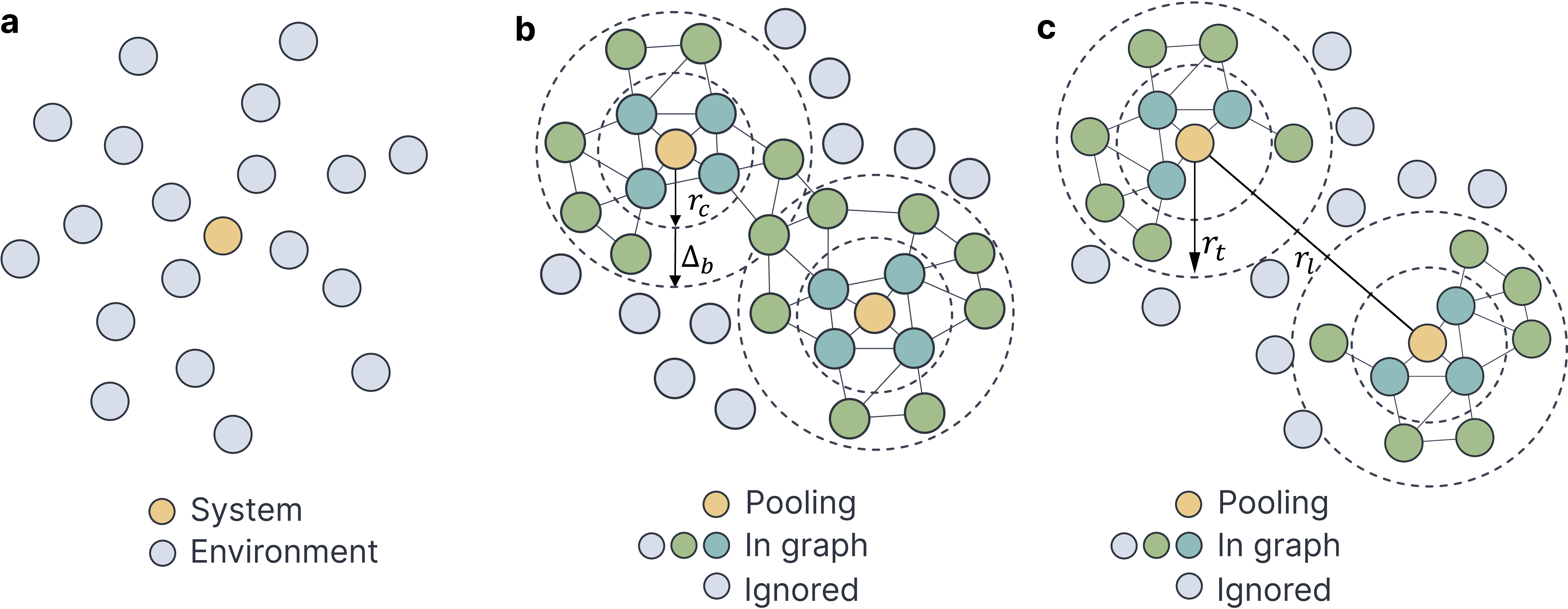}
\caption{\justifying \textbf{Dual-cutoff graph construction for dynamic molecular environments.}
(a) Atomic configuration partitioned into persistent system atoms and surrounding environment atoms.
(b) Local truncated graphs defined by a core radius $r_c$ and buffer width $\Delta_b$, giving an effective truncation radius $r_t=r_c+\Delta_b$; atoms outside the retained regions are ignored.
(c) The resulting dual-cutoff representation, where $r_t$ controls the truncation of the local environment, while the larger cutoff $r_l$ preserves interactions among persistent system atoms, allowing the graph to adapt dynamically as the surrounding environment evolves.}
\label{fig:extended-4}
\end{figure}

\begin{figure}[H]
\centering
\includegraphics[width=0.75\linewidth]{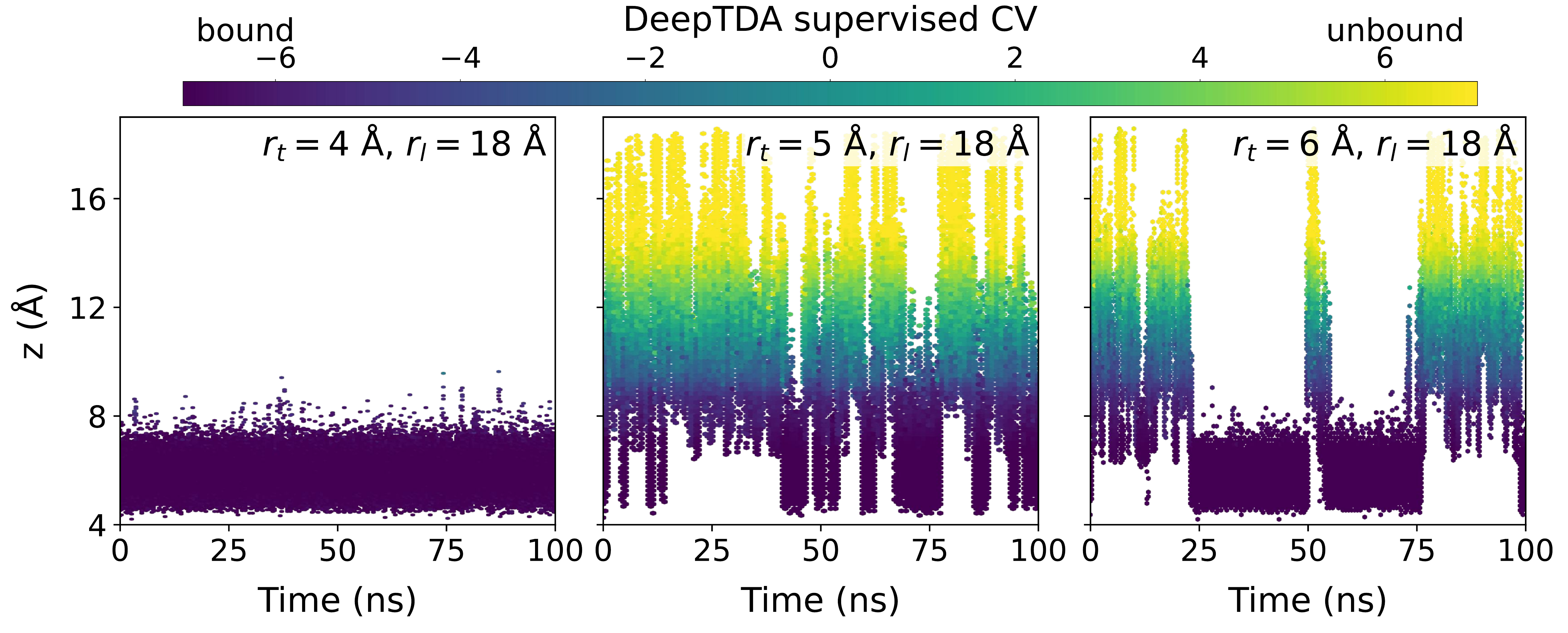}
\caption{\justifying \textbf{Solvent cutoff controls host--guest biased sampling.}
OPES-MetaD simulations of OAMe--G2 biased along DeepTDA CVs constructed using $r_l=18\,\text{\AA}$ and environment cutoffs of $r_t=4$, $5$, and $6\,\text{\AA}$. Points are colored according to the DeepTDA CV value. At $r_t=4\,\text{\AA}$, sampling remains largely confined to the bound state, whereas $r_t=5$ and $6\,\text{\AA}$ lead to repeated binding--unbinding transitions. These results show that including a sufficiently extended solvent environment is necessary to resolve the slow degrees of freedom associated with host--guest dissociation.}
\label{fig:extended-5}
\end{figure}
\end{onecolumngrid}

\clearpage
\begin{onecolumngrid}
\section*{Supplementary information}

\renewcommand{\thesection}{SUPPLEMENTARY SECTION \arabic{section}}
\setcounter{section}{0}

\captionsetup[figure]{format=plain}
\renewcommand{\figurename}{Supplementary Fig.}
\setcounter{figure}{0}

\renewcommand{\tablename}{Supplementary Table}

\renewcommand{\theequation}{S\arabic{equation}}
\setcounter{equation}{0}

\makeatletter
\setlength{\@fptop}{0pt}
\setlength{\@fpsep}{12pt plus 2pt minus 2pt}
\setlength{\@fpbot}{0pt plus 1fil}
\makeatother

\section{TRIPLE-WELL POTENTIAL - ADDITIONAL INFORMATION}
\subsection{Computational details}
\paragraphtitle{Simulation details}
The triple-well potential energy surface, \(U(x,y)\), is defined as a function of the Cartesian coordinates \(x\) and \(y\) as
\begin{equation}
\begin{aligned}
U(x,y) =\;& 3 e^{-x^2}
\left(
e^{-(y-\frac{1}{3})^2}
-
e^{-(y-\frac{5}{3})^2}
\right) \\
&- 5 e^{-y^2}
\left(
e^{-(x-1)^2}
+
e^{-(x+1)^2}
\right) \\
&+ 0.2 x^4
+ 0.2 \left(y-\frac{1}{3}\right)^4 .
\end{aligned}
\end{equation}

The simulation of the diffusion of an ideal particle of mass 1 has been performed using Langevin dynamics based on the Bussi-Parrinello algorithm \cite{bussi2007accurate} as implemented in the \texttt{ves\_md\_linearexpansion} \cite{valsson2014variational} module of \texttt{PLUMED} \cite{tribello2014plumed, plumed2019promoting}. The damping constant in the Langevin equation was set to 10/time-unit. The time unit was defined arbitrarily and corresponds to 200 timesteps. Unbiased simulations were performed at $k_BT=1.0$, whereas biased simulations were performed at the lower temperature of $k_BT=0.6$. All simulations were run for 20,000 time units. In the biased simulations, the OPES bias was updated every 500 steps, and the barrier parameter was set to 10~$k_{\mathrm{B}}T$.

\paragraphtitle{Training details}
The Cartesian coordinates $(x, y)$ of the diffusing particle were used as input features to a feed-forward neural network (FFNN) with architecture $[2, 50, 50, 5]$. The predictor was parameterized as a three-layer feedforward neural network, with each layer containing five neurons. A shifted softplus activation function was employed throughout all networks \cite{schutt2018schnet}. The contrastive loss was regularized with a coefficient of $1 \times 10^{-6}$. Model parameters were optimized using the ADAM optimizer with a learning rate of $1 \times 10^{-3}$. Training was performed for 500 epochs. For the numerical stability benchmark, the lag time was set to $\tau = 1$ time unit, corresponding to 200 integration steps.

\subsection{Additional results}
\begin{figure}[H]
\centering
\includegraphics[width=0.8\linewidth]{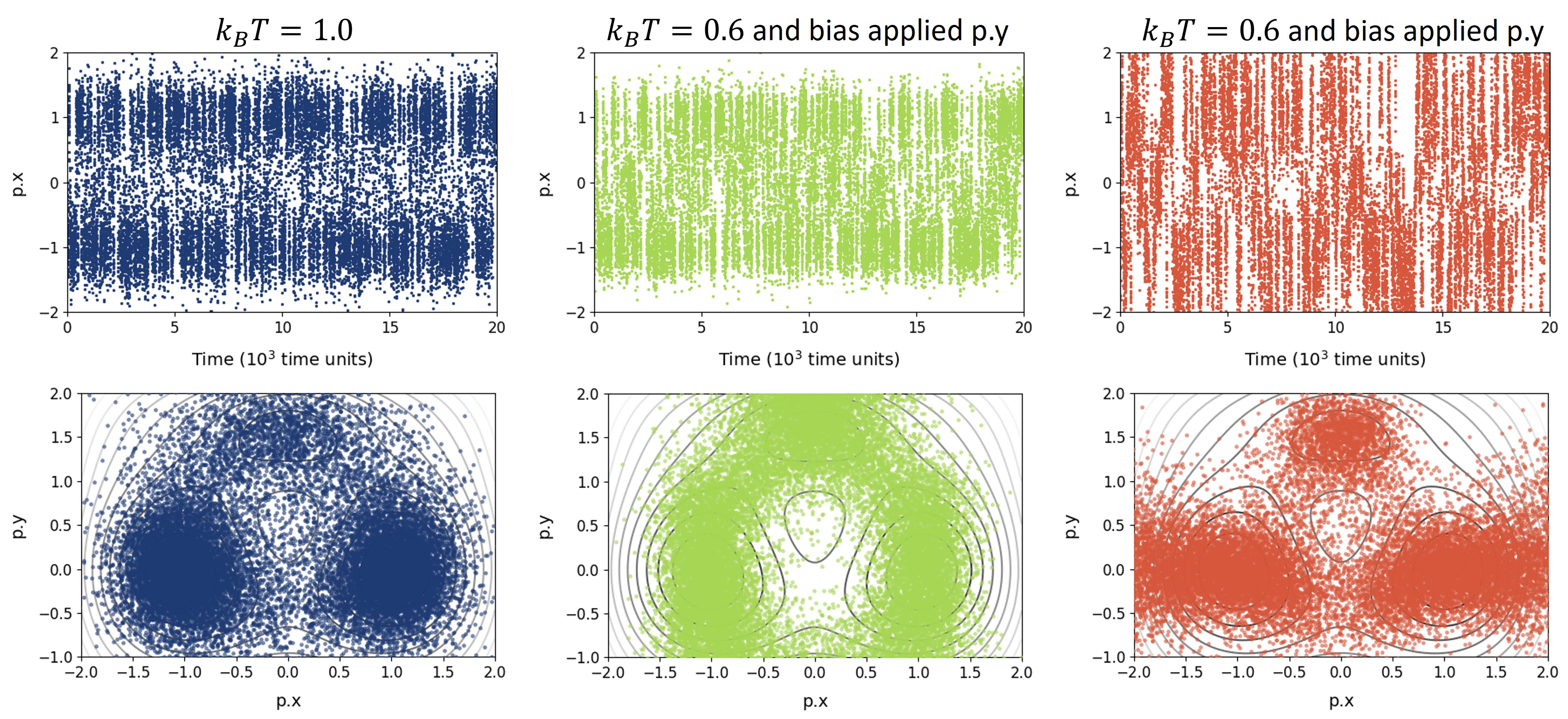}  
\caption{The distributions obtained from different trajectories and the time evolution of $p_x$. The blue and green curves correspond to unbiased simulations at $k_B T = 0.6$ and $k_B T = 1.0$, respectively, while the red curve corresponds to a biased simulation at $k_B T = 0.6$ with a bias applied along the $y$ coordinate. }
\label{fig:Tri_dataset}
\end{figure}

\begin{figure}[H]
\centering
\includegraphics[width=0.8\linewidth]{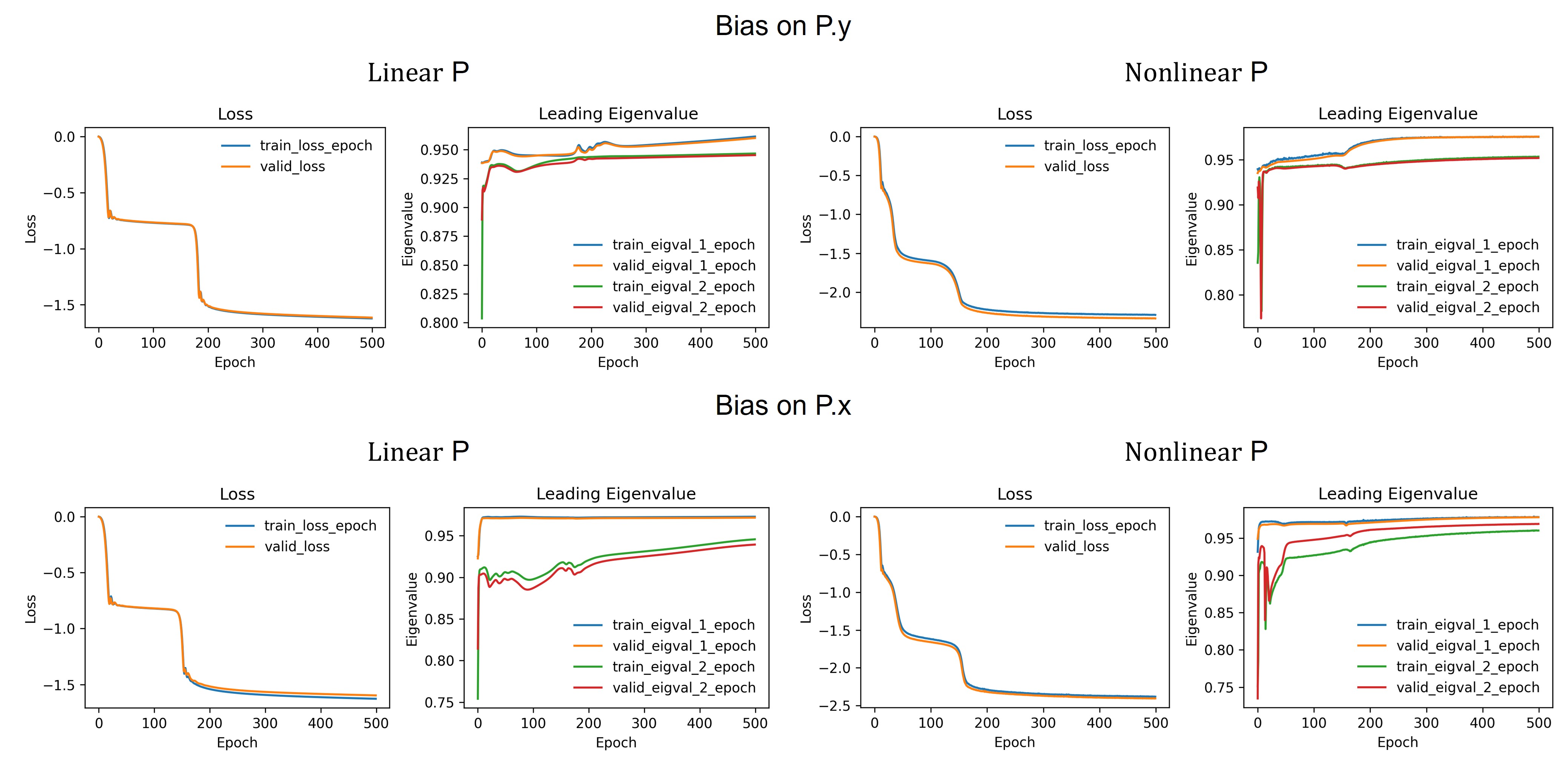}  
\caption{Training stability of SelfTICA on biased triple-well trajectories. Results are shown for trajectories biased along the $y$-coordinate (top row) and the $x$-coordinate (bottom row), using either a linear or a nonlinear predictor. Each block reports the training and validation loss curves together with the two leading TICA eigenvalues estimated from the learned representations.}
\label{fig:Tri_loss}
\end{figure}

\begin{figure}[H]
\centering
\includegraphics[width=0.8\linewidth]{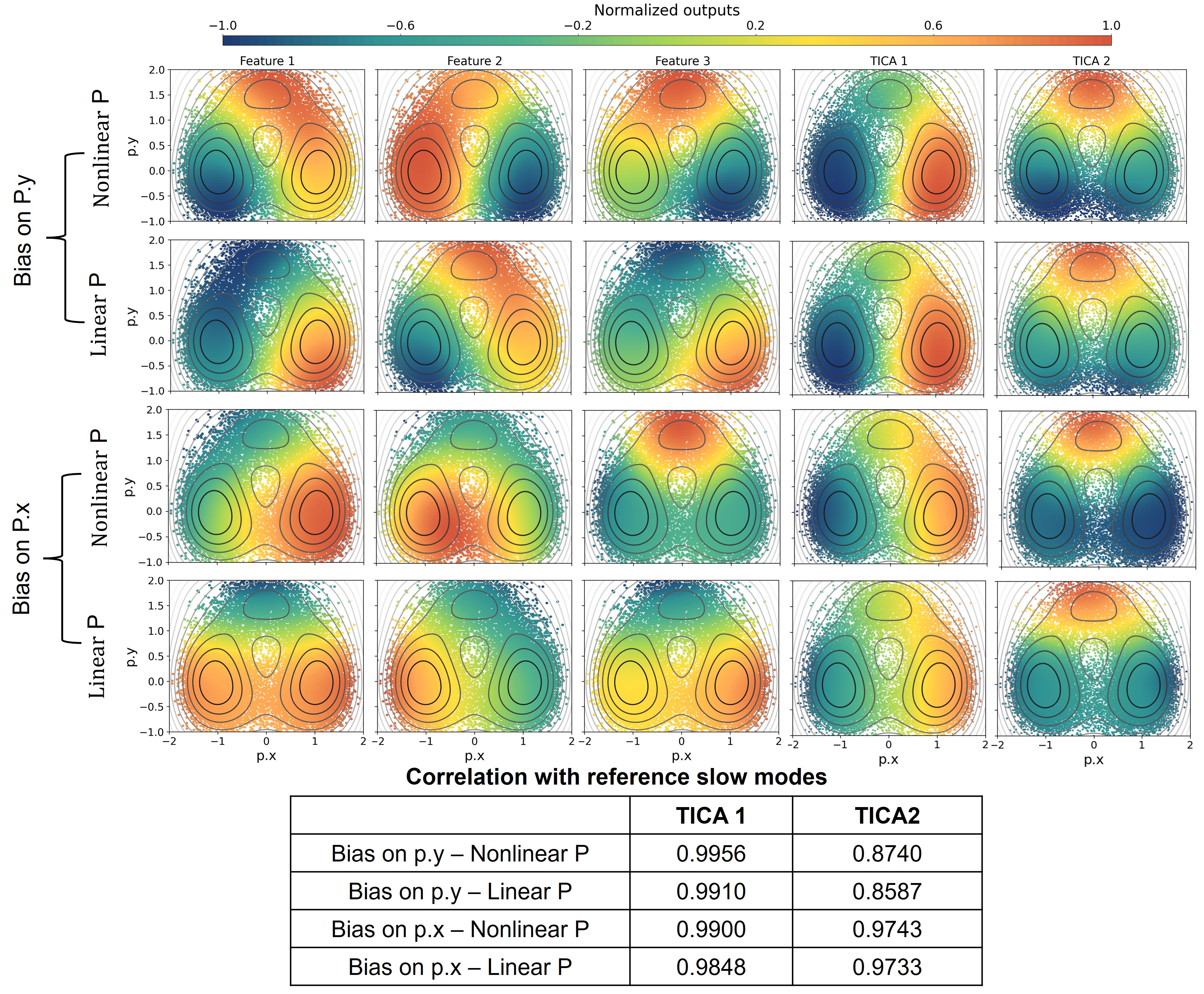}  
\caption{Encoder features learned from trajectories biased along the $y$-coordinate, together with the corresponding TICA projections obtained using linear and nonlinear predictors.}
\label{fig:Tri_features}
\end{figure}

\begin{figure}[H]
\centering
\includegraphics[width=0.4\linewidth]{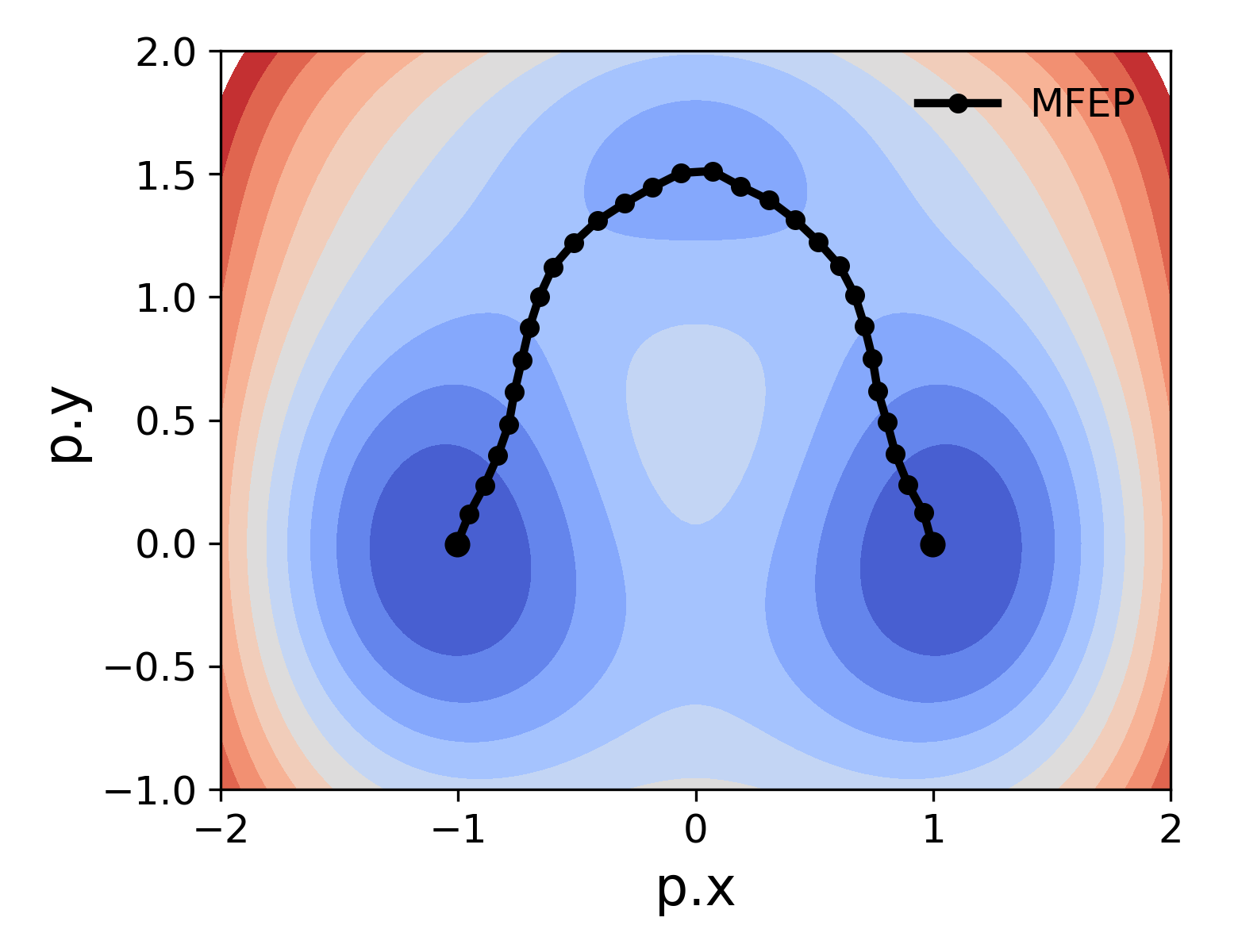}  
\caption{Minimum free energy path (MFEP) projected onto the two-dimensional configuration space. The black curve with markers denotes the MFEP connecting the two metastable basins on the potential energy surface.}
\label{fig:Tri_mfep}
\end{figure}

\begin{figure}[H]
\centering
\includegraphics[width=0.8\linewidth]{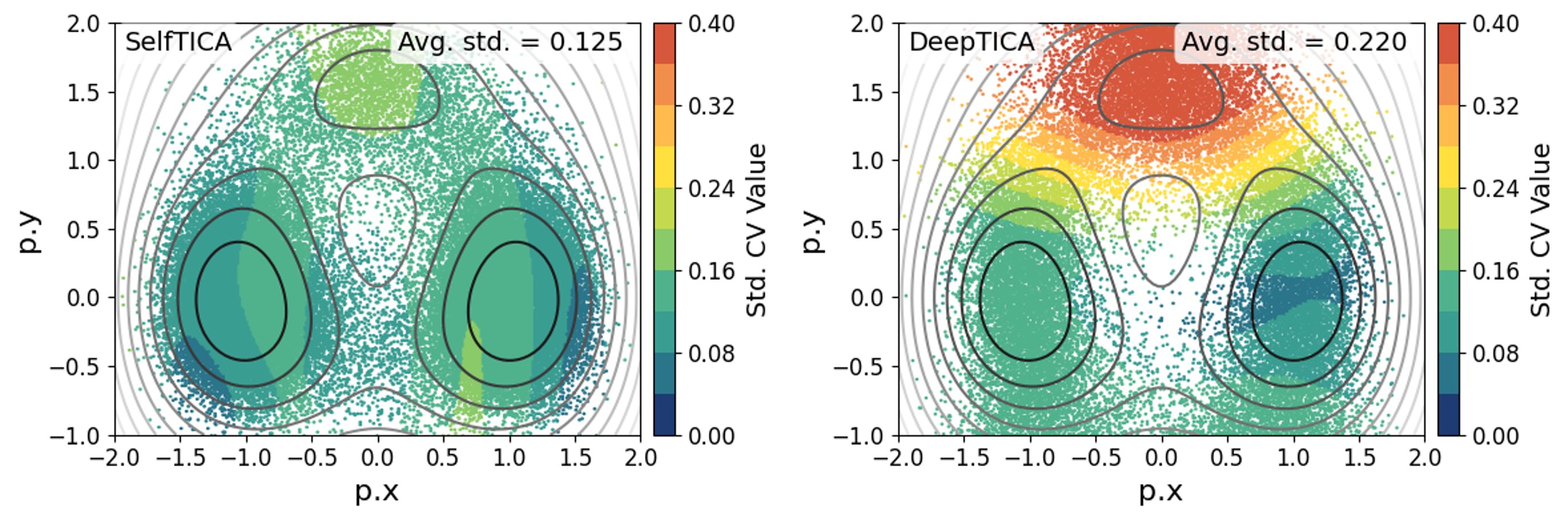}  
\caption{Projection of the standard deviation across 25 independently trained SelfTICA and DeepTICA models onto the configuration space.}
\label{fig:Tri_std}
\end{figure}
\clearpage

\section{ALANINE DIPEPTIDE - ADDITIONAL INFORMATION}
\subsection{Computational details}
\paragraphtitle{Simulation details}
All simulations of alanine dipeptide (Ace-Ala-Nme) in vacuum were performed using the \texttt{GROMACS} v2022.5\cite{abraham2015gromacs} molecular dynamics engine patched with \texttt{PLUMED}\cite{tribello2014plumed,plumed2019promoting}, employing the Amber99SB force field \cite{salomon2013overview} and a 2~fs integration timestep. Langevin dynamics \cite{bussi2007accurate} was used for thermostatting at a temperature of 300~K, with a damping coefficient given by $\gamma_i = m_i / (\tau - t)$, where $\tau - t = 0.05$~ps.

For the trial simulation, an OPES multithermal simulation was carried out over a temperature range of 300--600~K for a total duration of 1000~ns. The first 100~ns of the trajectory were discarded to ensure that the bias had reached a quasi-stationary regime. Frames from 100--120~ns, 100--200~ns, and 100--600~ns were then used to construct the training datasets, denoted as dataset 1, dataset 2, and dataset 3, respectively. The remaining 600--1000~ns segment was held out as the test set. To assess the stability of the learned CVs, OPES-MetaD biased simulations driven by each trained SelfTICA or DeepTICA CV were performed for 20~ns. In these simulations, the OPES bias was updated every 500 steps, and the barrier parameter was set to 40~kJ~mol$^{-1}$.

To evaluate implied timescales, additional unbiased simulations were performed at 450, 500, and 600~K, with trajectory lengths of 500, 200, and 100~ns, respectively. These trajectories were not used for training and served only to assess whether the learned representation preserves the temperature-dependent relaxation times of the slow process. For each temperature, the trained SelfTICA encoder was kept fixed, TICA was applied to the learned features at different lag times, and the implied timescales were computed from the corresponding TICA eigenvalues.

\paragraphtitle{Training details}
For the FFNN-based models, we used 45 pairwise heavy-atom distances as input features to a feed-forward neural network with architecture $[45,30,30,5]$. The predictor was parameterized as a three-layer feedforward neural network, with each layer containing five neurons. Rectified linear unit (ReLU) activations were employed in all hidden layers.

For the GNN-based models, all heavy atoms in the system were treated as reactive atoms, and a cutoff of $r_c = 10$~\AA\ was applied. The GNN consisted of three message-passing layers, each using 16 Gaussian basis functions and 32 filters, with 64 hidden channels. The final output feature dimension was set to 8. Message aggregation was performed using a minimum-value operation.

The contrastive loss was regularized with a coefficient of $1 \times 10^{-6}$. Model parameters were optimized using the Adam optimizer with a learning rate of $1 \times 10^{-3}$. Training was carried out for 500, 200, and 100 epochs on Dataset 1, Dataset 2, and Dataset 3, respectively. For the numerical stability benchmark, the lag time was set to $\tau = 10$.

\subsection{Additional results}
\begin{figure}[H]
\centering
\includegraphics[width=0.7\linewidth]{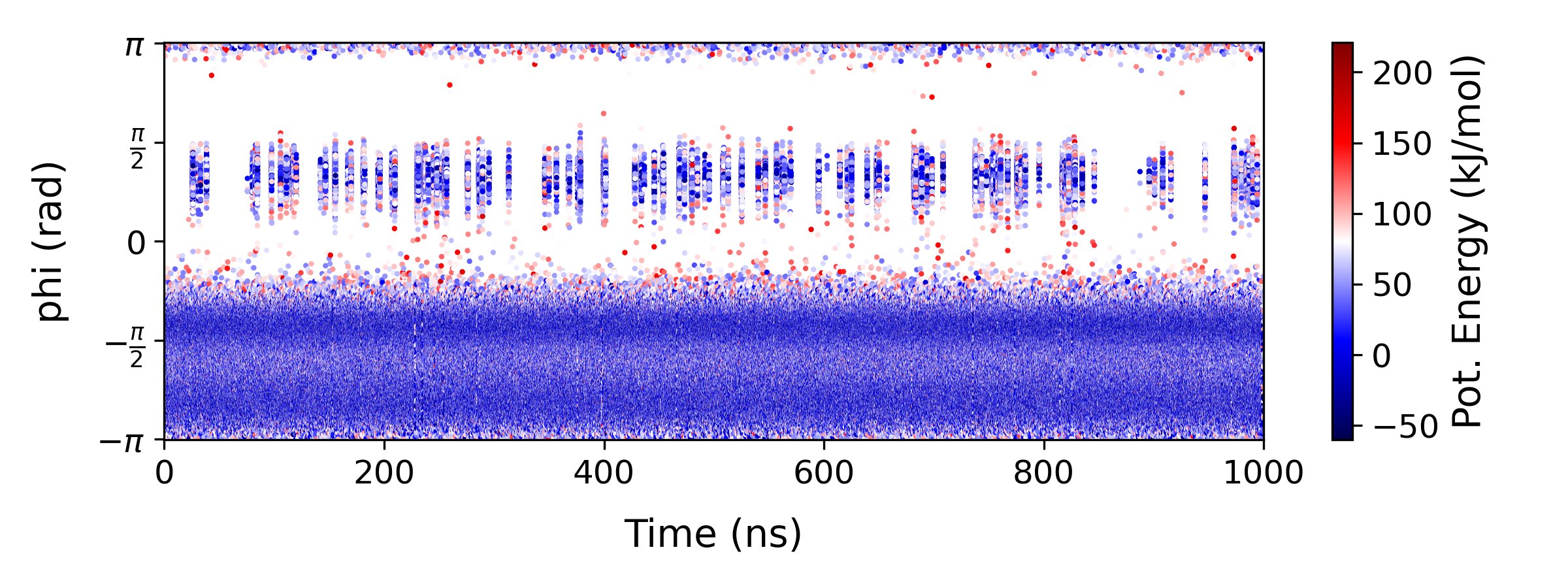}  
\caption{The time evolution of the $\phi$ angle in the initial multithermal run. The points are colored according to their potential energy.}
\label{fig:ala2_multi}
\end{figure}

\begin{figure}[H]
\centering
\includegraphics[width=0.8\linewidth]{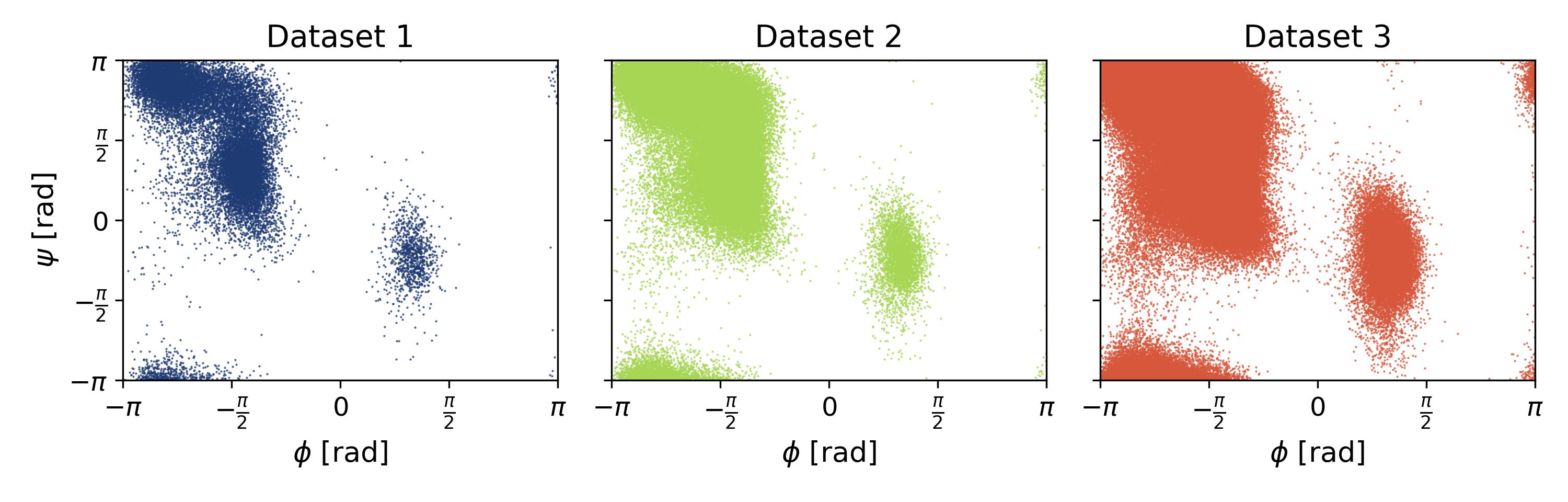}  
\caption{Distribution of alanine dipeptide training datasets with increasing trajectory lengths. Scatter plots show the sampled configurations projected onto the $\phi$--$\psi$ dihedral-angle space. Dataset 1, Dataset 2, and Dataset 3 were constructed from frames in the 100--120 ns, 100--200 ns, and 100--600 ns time windows, respectively.}
\label{fig:ala2_dataset}
\end{figure}

\begin{figure}[H]
\centering
\includegraphics[width=0.8\linewidth]{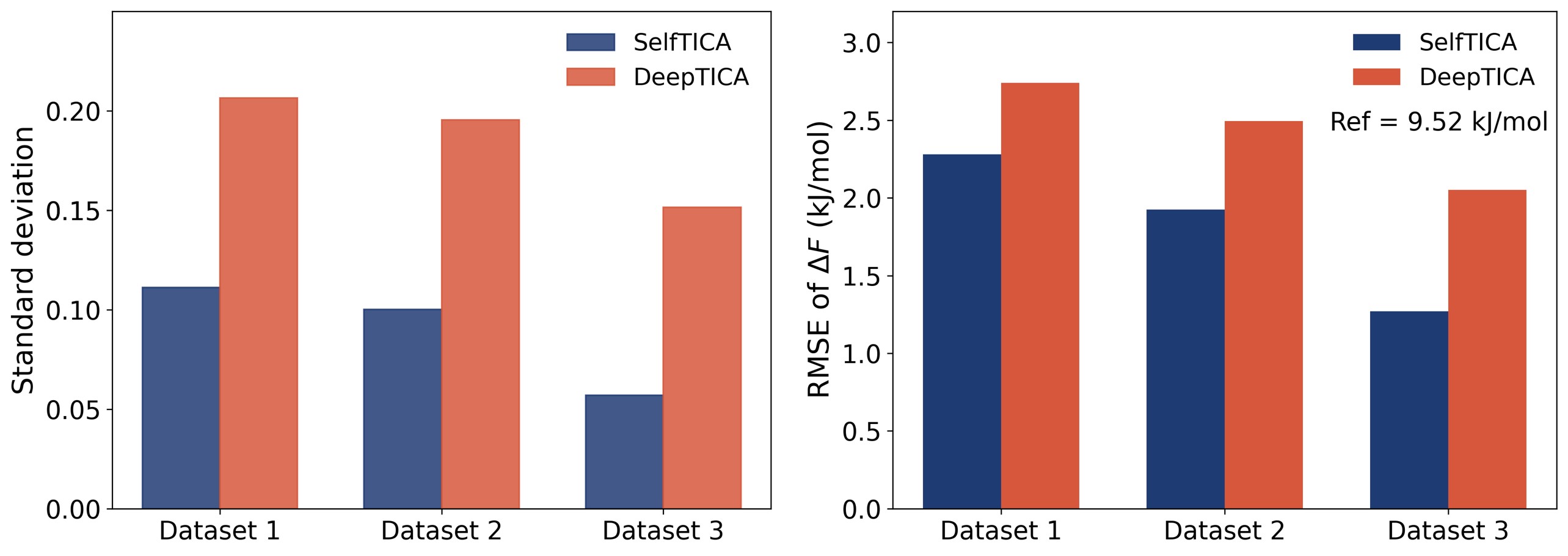}  
\caption{Standard deviation and RMSE of free-energy differences $\Delta F$ (in $k_{\mathrm B}T$), obtained from GNN-based SelfTICA and DeepTICA models trained on datasets of different lengths and evaluated over 10 OPES simulations.}
\label{fig:ala2_gnn}
\end{figure}



\begin{figure}[H]
\centering
\includegraphics[width=1.0\linewidth]{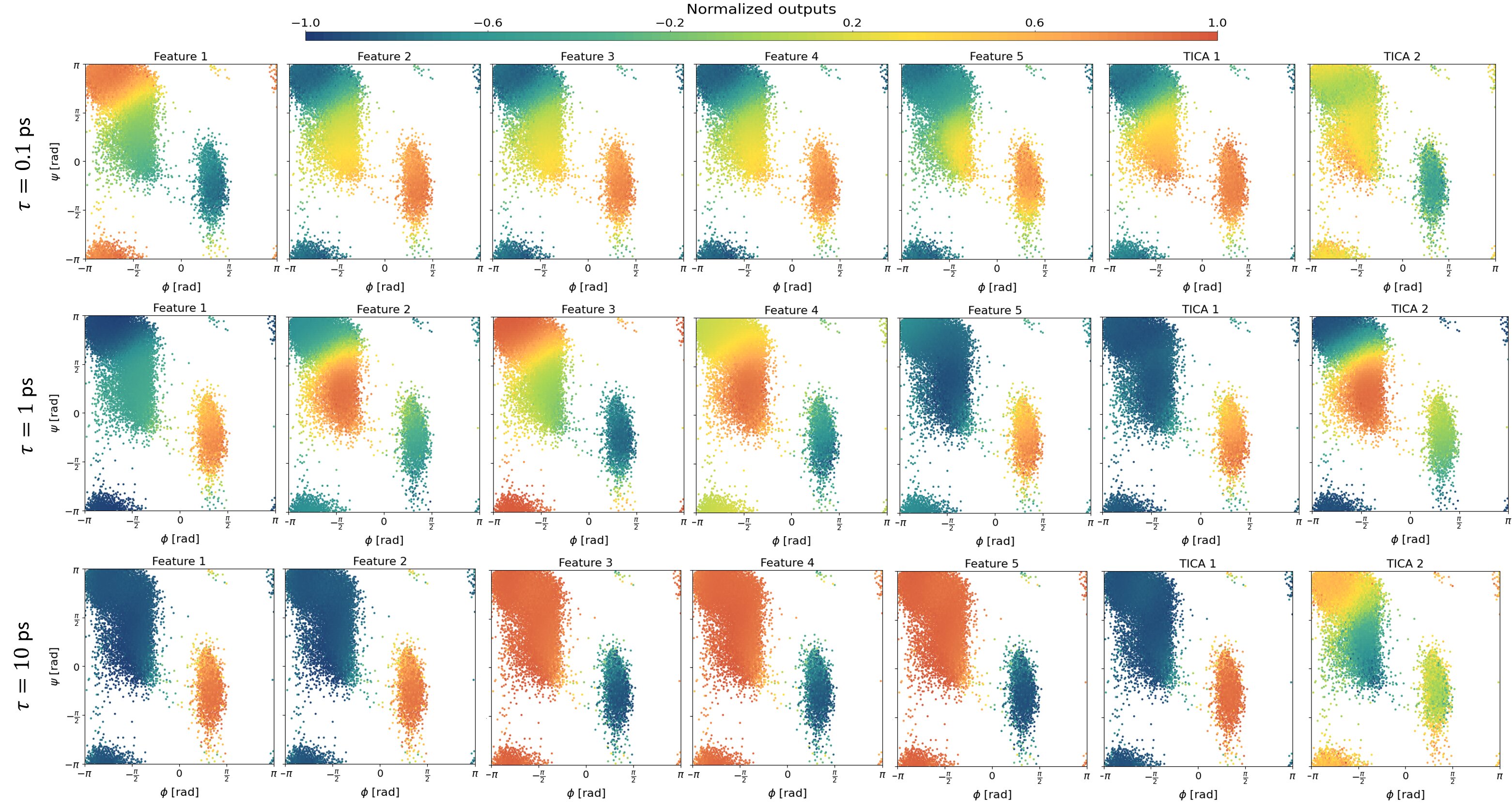}  
\caption{Encoder output features and corresponding TICA components learned by SelfTICA models trained at different lag times.}
\label{fig:ala2_feature}
\end{figure}
\clearpage

\section{Chignolin mini-protein folding - ADDITIONAL INFORMATION}
\subsection{Computational details}
\paragraphtitle{Simulation details}
To investigate the folding and unfolding behavior of chignolin (CLN025; peptide sequence Tyr–Tyr–Asp–Pro–Glu–Thr–Gly–Thr–Trp–Tyr) in explicit solvent, we carried out molecular dynamics simulations using \texttt{GROMACS} v2024.5 \cite{abraham2015gromacs} patched with \texttt{PLUMED}\cite{tribello2014plumed,plumed2019promoting}, employing the $\text{CHARMM22}^{*}$ \cite{piana2011robust} force field. The solvent environment was modeled using the CHARMM TIP3P \cite{mackerell1998all} water model. This setup is consistent with previous long-timescale unbiased simulations on this system\cite{lindorff2011fast}, enabling direct comparison of results.

All simulations were conducted in the NVT ensemble at 340 K, using a 2 fs integration time step. The Asp and Glu residues, as well as the N- and C-termini, were modeled in their charged states. The simulation box contained 1,907 water molecules and two sodium ions to ensure charge neutrality. Bond constraints involving hydrogen atoms were enforced using the LINCS \cite{hess1997lincs} algorithm, and long-range electrostatic interactions were computed via the particle mesh Ewald (PME) \cite{darden1993particle} method, with a 1 nm cutoff for all nonbonded interactions.

For the trial simulation, a 1000~ns OPES-explore simulation was carried out at 340~K using the C$_\alpha$ RMSD as a deliberately simple preliminary CV. For the production simulations, OPES-explore was applied to bias the first SelfTICA CV. In all OPES-explore simulations, the bias was updated every 500 steps and the barrier parameter was set to 20~kJ~mol$^{-1}$.

\paragraphtitle{Training details}
To train the models, we used 210 interatomic distances as input descriptors. The DeepTDA model was implemented as an FFNN with architecture $[210,50,50,1]$, using target centers of $[-7.0,7.0]$ and target sigmas of $[0.2,0.2]$. For a fair comparison, DeepTICA and SelfTICA used the same encoder backbone, defined as an FFNN with architecture $[210,50,50,5]$. In SelfTICA, the predictor was parameterized as a three-layer feedforward neural network, with each layer containing five neurons. A lag time of $\tau = 50$ ps was used for both SelfTICA and DeepTICA training. ReLU activations were employed throughout all networks. The contrastive loss was regularized with a coefficient of $1 \times 10^{-6}$.

Model parameters were optimized using the Adam optimizer with a learning rate of $1 \times 10^{-3}$. The SelfTICA and DeepTICA models were trained for 100 epochs, whereas the DeepTDA model was trained for 500 epochs to ensure convergence.

\subsection{Additional results}

\begin{figure}[H]
\centering
\includegraphics[width=0.7\linewidth]{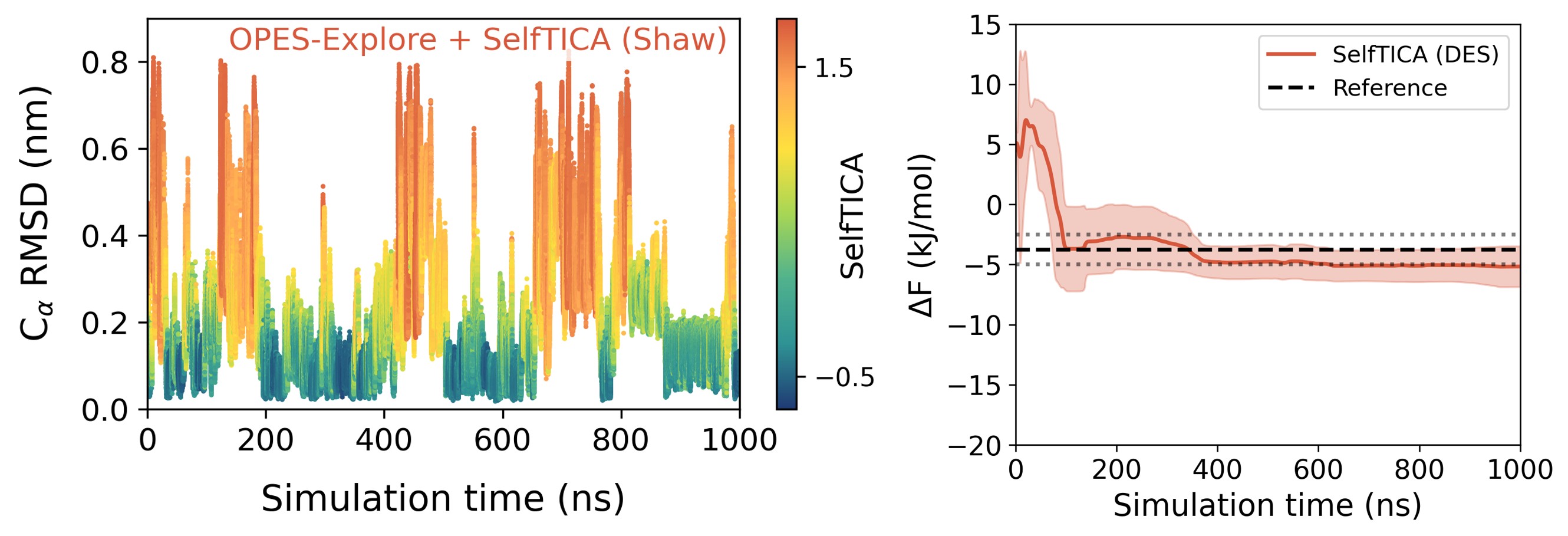}  
\caption{Reference enhanced-sampling performance of SelfTICA trained on the long unbiased D. E. Shaw trajectory. Left, time evolution of the C$_\alpha$ RMSD in an OPES-explore simulation biased along the resulting SelfTICA CV, with points colored by the CV value. Right, convergence of the folding free-energy difference $\Delta F$ toward the reference value.}
\label{fig:chig_tica_shaw}
\end{figure}

\clearpage

\section{Calixarene host-guest systems - ADDITIONAL INFORMATION}
\subsection{Computational details}
\paragraphtitle{Simulation details}
We performed all simulations using \texttt{GROMACS}~v2024.5~\cite{abraham2015gromacs} in combination with the \texttt{PLUMED}~\cite{tribello2014plumed,plumed2019promoting} plugin. The system was described using the GAFF~\cite{wang2004development} force field with RESP charges, and the solvent was modeled with the TIP3P water model. Simulations were conducted at a temperature of 300~K using a velocity-rescaling thermostat~\cite{bussi2007canonical} with a time constant of 0.1~ps, and an integration time step of 2~fs was employed. The simulation box was cubic with a side length of 40.27~\AA, containing 2,100 water molecules, the OAMe host, and the selected guest molecule. Sodium ions were added to neutralize the system. Following the standard SAMPL5 host--guest setup, a virtual atom V1 was defined at the center of the OAMe host and used as the reference point for the binding coordinate. At each simulation step, the coordinates were aligned so that the vertical axis of the box coincided with the binding axis $h$, and the box was centered on V1.

To initiate the trial simulations, we first carried out 30~ns unbiased MD simulations in both the bound and unbound states to train the DeepTDA CVs. We then performed 100~ns OPES-MetaD simulations using the learned DeepTDA CVs. For production, we carried out 250~ns OPES-MetaD simulations using the first SelfTICA CV, while retaining the final static DeepTDA bias to preserve the broad configurational coverage generated during the initial exploration. In all OPES-MetaD simulations, the bias was updated every 500 steps, and the barrier parameter was set to 50~kJ~mol$^{-1}$.

\paragraphtitle{The funnel restraint}
Our simulations adopt the funnel-shaped restraint introduced by Limongelli et al.~\cite{limongelli2013funnel} and used in previous studies~\cite{rizzi2021role,bhakat2017resolving}. The restraint confines the ligand within a cylindrical volume above the binding site, thereby restricting the accessible space of the unbound state (U), while widening near the binding pocket so as not to perturb the binding process. After aligning the system with \texttt{PLUMED} to a reference configuration in which the binding axis coincides with the vertical ($z$) direction, we define $z$ as the projection of the geometric center of the ligand carbon atoms onto this axis. For $z > 10$~\AA, the funnel surface is a cylinder of radius $R_\mathrm{cyl} = 2$~\AA. For $z < 10$~\AA, the funnel opens with a 45$^\circ$ angle and is described by $r = 12 - z$, where $r$ is the radial distance from the funnel axis. When the ligand crosses the funnel surface, a harmonic restraining force $F = -k_\mathrm{F} x$ is applied, where $x$ is the displacement from the surface and $k_\mathrm{F} = 20$~kJ~mol$^{-1}$~\AA$^{-2}$. An additional harmonic restraint along $z$ prevents the ligand from drifting too far from the host and reaching the upper boundary of the simulation box, with $F = -k_\mathrm{U}(z - 18)$ for $z > 18$~\AA\ and $k_\mathrm{U} = 40$~kJ~mol$^{-1}$~\AA$^{-2}$.

Because the funnel restraint limits the volume accessible to the unbound ligand, the free energy difference extracted from enhanced sampling simulations requires a standard-state correction:
\begin{equation}
\Delta G = -\frac{1}{\beta} \log \left( C_0 \pi R_{\mathrm{cyl}}^2 \int_{\mathcal{B}} \mathrm{d}z , \exp \left[ -\beta \left( W(z) - W_{\mathrm{u}} \right) \right] \right),
\end{equation}
where $\beta = 1/(k_B T)$, $C_0 = 1/1660\ \mathrm{\AA}^{-3}$ is the standard concentration, $W(z)$ is the free energy along the funnel axis, and $W_\mathrm{u}$ is the reference free energy of the unbound state. We define $W_\mathrm{u}$ as the average free energy in the interval $16\ \mathrm{\AA} < z < 18\ \mathrm{\AA}$, and compute the integral over the bound-state region $\mathcal{B}$, defined as $3\ \mathrm{\AA} < z < 8\ \mathrm{\AA}$.

\paragraphtitle{Training details}
We employed a GNN to construct collective variables for the calixarene system. Specifically, 7 atoms from the guest molecule and 11 atoms from the host backbone were selected as reactive atoms, while nearby water oxygen atoms were treated as environment atoms. Models were trained using short-range cutoff radii of $r_c = 3, 4, 5$~\AA, together with an additional buffer of $\Delta b = 1$~\AA, corresponding to effective cutoffs of $r_t = 4, 5, 6$~\AA. A fixed long-range cutoff of $r_l = 18$~\AA\ was applied to all reactive atoms. As a control, we also trained a host--guest-only model in which the graph representation included only the selected reactive atoms, excluded water oxygen atoms, and used a cutoff distance of $18~\text{\AA}$.

For the DeepTDA models, the GNN consisted of two message-passing layers with 16 Gaussian basis functions and 16 filters per layer, each employing 20 hidden channels, and using target centers of $[-7.0,7.0]$ and target sigmas of $[0.2,0.2]$. For the SelfTICA models, we employed a SchNet architecture with two message-passing layers, 12 Gaussian basis functions, and 12 filters per layer, each with 12 hidden channels, followed by an output feature dimension of 4. Message aggregation was performed using an attention mechanism. The predictor was parameterized as a three-layer feedforward neural network, with each layer containing four neurons.

The contrastive loss was regularized with a coefficient of $1 \times 10^{-6}$. Model parameters were optimized using the Adam optimizer with a learning rate of $1 \times 10^{-3}$. The SelfTICA models were trained for 100 epochs, whereas the DeepTDA models were trained for 500 epochs to ensure convergence. The lag time used to construct the time-lagged training dataset was set to $\tau = 1$ ps.

\subsection{Additional results}
\begin{figure}[H]
\centering
\includegraphics[width=0.5\linewidth]{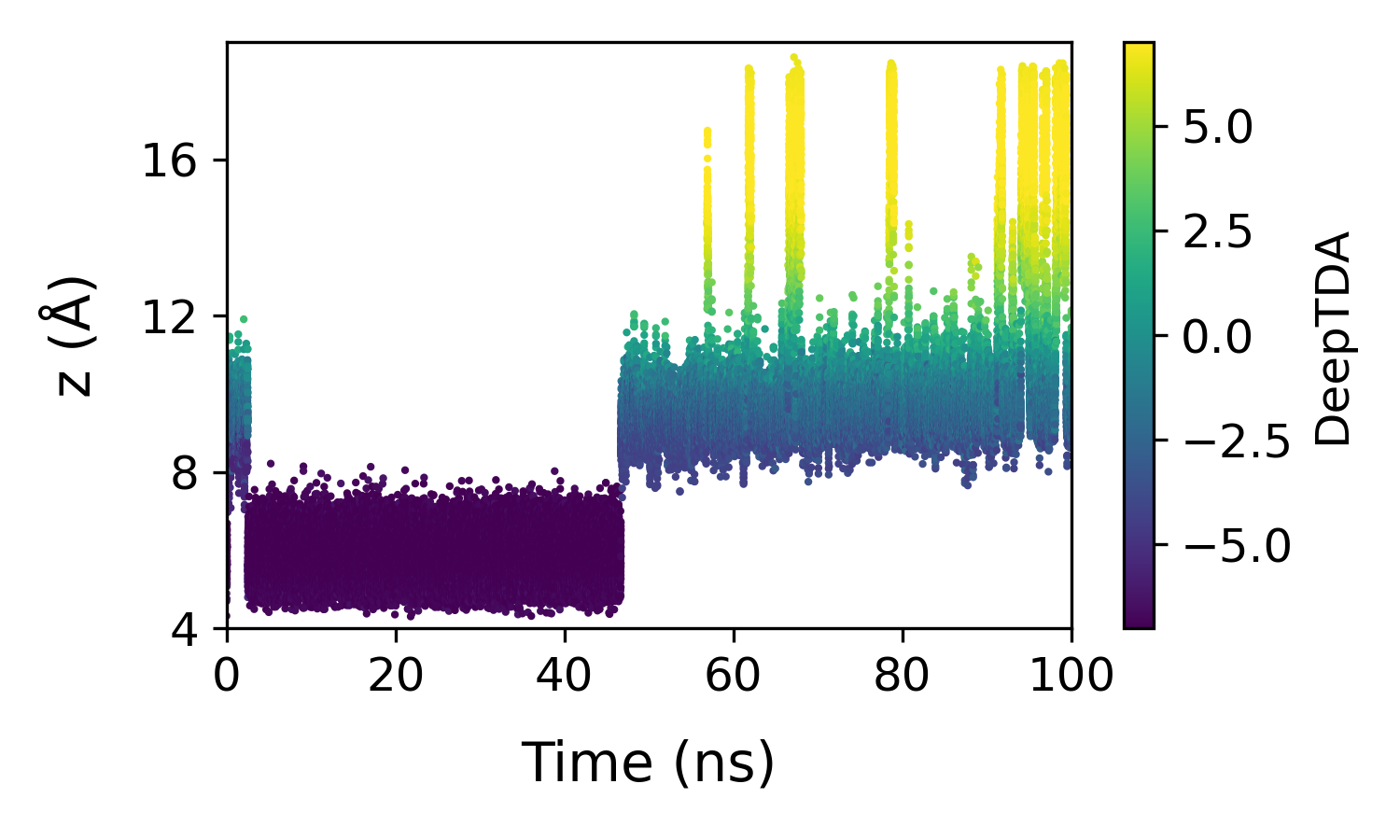}  
\caption{OPES simulation driven by a graph-based DeepTDA CV. 
The ligand position $z$ is plotted over time and colored by the DeepTDA CV value. 
Only selected host--guest atoms were included in the graph, with a cutoff distance of 18~\AA{}.
}
\label{fig:calixarene_noenv}
\end{figure}

\begin{figure}[H]
\centering
\includegraphics[width=0.8\linewidth]{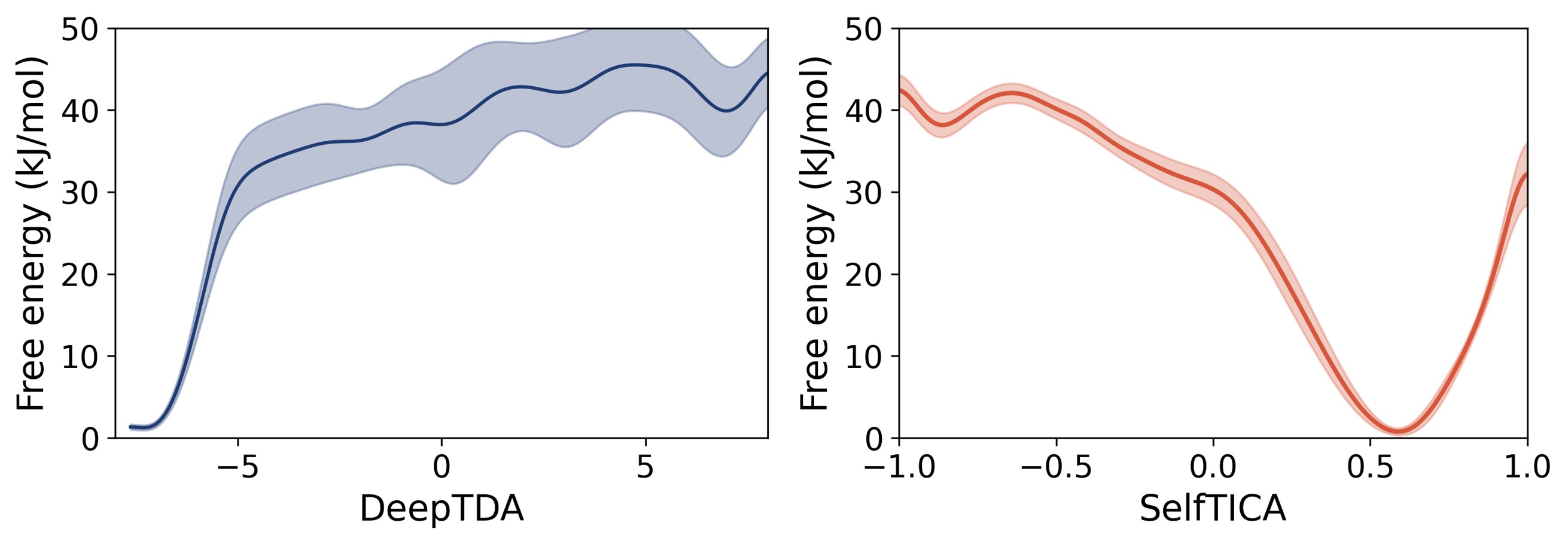}  
\caption{Free energy profiles for the SelfTICA simulation. The average estimates from three independent simulations are reported as a green solid line, whereas the uncertainty, computed as the standard deviation over three replicas, is depicted as a shaded green region. The reference values are provided as gray dashed lines, and the 0.5 $k_BT$ interval around the reference is marked by gray dotted lines.}
\label{fig:calixarene_fes}
\end{figure}

\clearpage

\section{Catalytic Dissociation of $\mathrm{N_2}$ on Fe(111) Surfaces - ADDITIONAL INFORMATION}

\subsection{Computational details}
\paragraphtitle{Simulation details}
Simulations of the catalytic dissociation of $\mathrm{N_2}$ on the $\mathrm{Fe(111)}$ surface were carried out using the \texttt{LAMMPS} software \cite{thompson2022lammps}, patched with MACE \cite{batatia2022mace} and \texttt{PLUMED} \cite{tribello2014plumed}. The system consisted of a total of 194 atoms, including 192 Fe atoms forming the slab and 2 N atoms comprising the $\mathrm{N_2}$ molecule. The simulation box dimensions were $16.235 \times 14.060 \times 39.113$~\AA$^3$. The interatomic potential was described using an MACE model \cite{batatia2022mace} trained via an active learning procedure \cite{perego2024data}, enabling an accurate representation of the potential energy surface across the relevant configurational space. All simulations were performed in the NVT ensemble with an integration time step of $0.5\,\mathrm{fs}$. The temperature was controlled using a stochastic velocity-rescaling thermostat \cite{bussi2007canonical} with a coupling time constant of $100\,\mathrm{fs}$.

In the simulation setup, the bottom two layers of the slab were kept fixed to impose boundary conditions mimicking a semi-infinite surface. Periodic boundary conditions were applied in the $x$ and $y$ directions, while along the $z$ direction, a reflecting wall was placed 10.4~\AA{} above the surface. During sampling, an upper wall was applied to the nitrogen--nitrogen distance, $d_{\mathrm{NN}}$, at 2.2~\AA{} to restrict the exploration to the relevant dissociation region.

During sampling, the iron--nitrogen coordination number was computed as a continuous and differentiable switching function,
\begin{equation}
C_{N,\mathrm{Fe}} = \sum_{i \in {N}} \sum_{j \in {\mathrm{Fe}}}
\frac{1 - \left( r_{ij}/r_0 \right)^n}
{1 - \left( r_{ij}/r_0 \right)^m},
\end{equation}
where $r_{ij}$ is the distance between nitrogen atom $i$ and iron atom $j$. The sums run over the two nitrogen atoms of $\mathrm{N_2}$ and all Fe atoms included in the coordination-number definition. The switching parameters were set to $r_0 = 2.5,\mathrm{\AA}$, $n = 6$, and $m = 12$, so that Fe--N contacts shorter than $r_0$ contribute strongly, whereas more distant pairs contribute smoothly toward zero.

For all OPES simulations of the $\mathrm{N_2}$ dissociation system, including both the initial OPES-MetaD run and the subsequent OPES-explore simulations, the bias was updated every 500 steps, and the barrier parameter was set to 82~kJ~mol$^{-1}$.

\paragraphtitle{Training details}
We employed a GNN to train the MLCV for this system. All 194 atoms were represented as nodes in the graph, and interatomic edges were constructed using a cutoff radius of $r_c = 5$~\AA, enabling the model to capture interactions spanning approximately two layers of Fe atoms.

The GNN architecture used within SelfTICA consisted of two message-passing layers, each using 16 Gaussian radial basis functions and 20 filters, with 20 hidden channels per layer, followed by a latent feature representation of dimension 8. Message aggregation was performed using mean pooling. The predictor was parameterized as a three-layer feedforward neural network, with each layer containing eight neurons.

The contrastive loss was regularized with a coefficient of $1 \times 10^{-6}$. Model parameters were optimized using the Adam optimizer with a learning rate of $1 \times 10^{-3}$. The SelfTICA model was trained for 100 epochs. The lag time used to construct the time-lagged training dataset was set to $\tau = 0.05$ ps.

\subsection{Additional results}
\begin{figure}[H]
\centering
\includegraphics[width=0.4\linewidth]{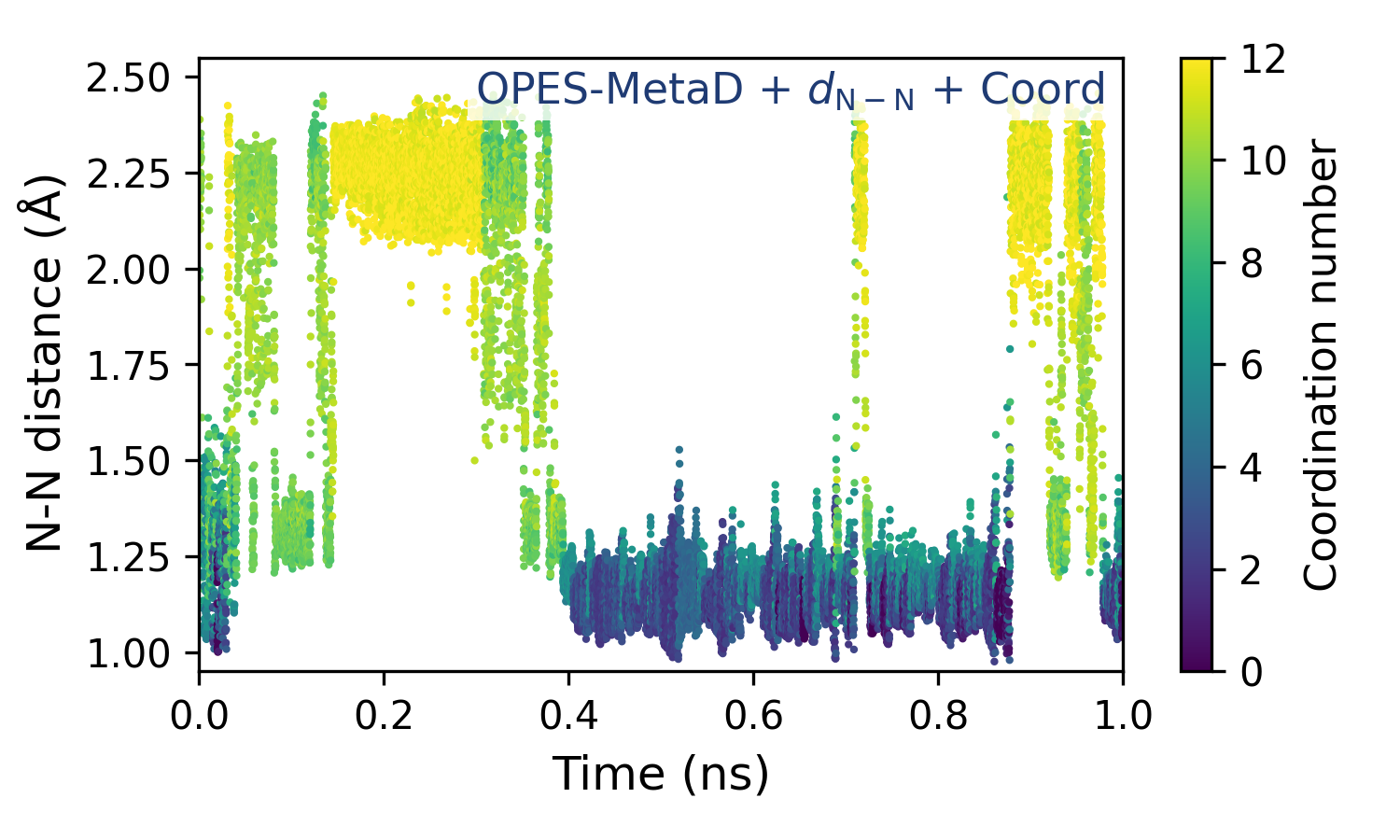}  
\caption{Time evolution of the $\mathrm{N\!-\!N}$ distance from OPES-MetaD simulation driven by the $\mathrm{N\!-\!N}$ distance and coordination number. Data points are colored according to the coordination number.}
\label{fig:fen2_initial}
\end{figure}

\begin{figure}[H]
\centering
\includegraphics[width=0.4\linewidth]{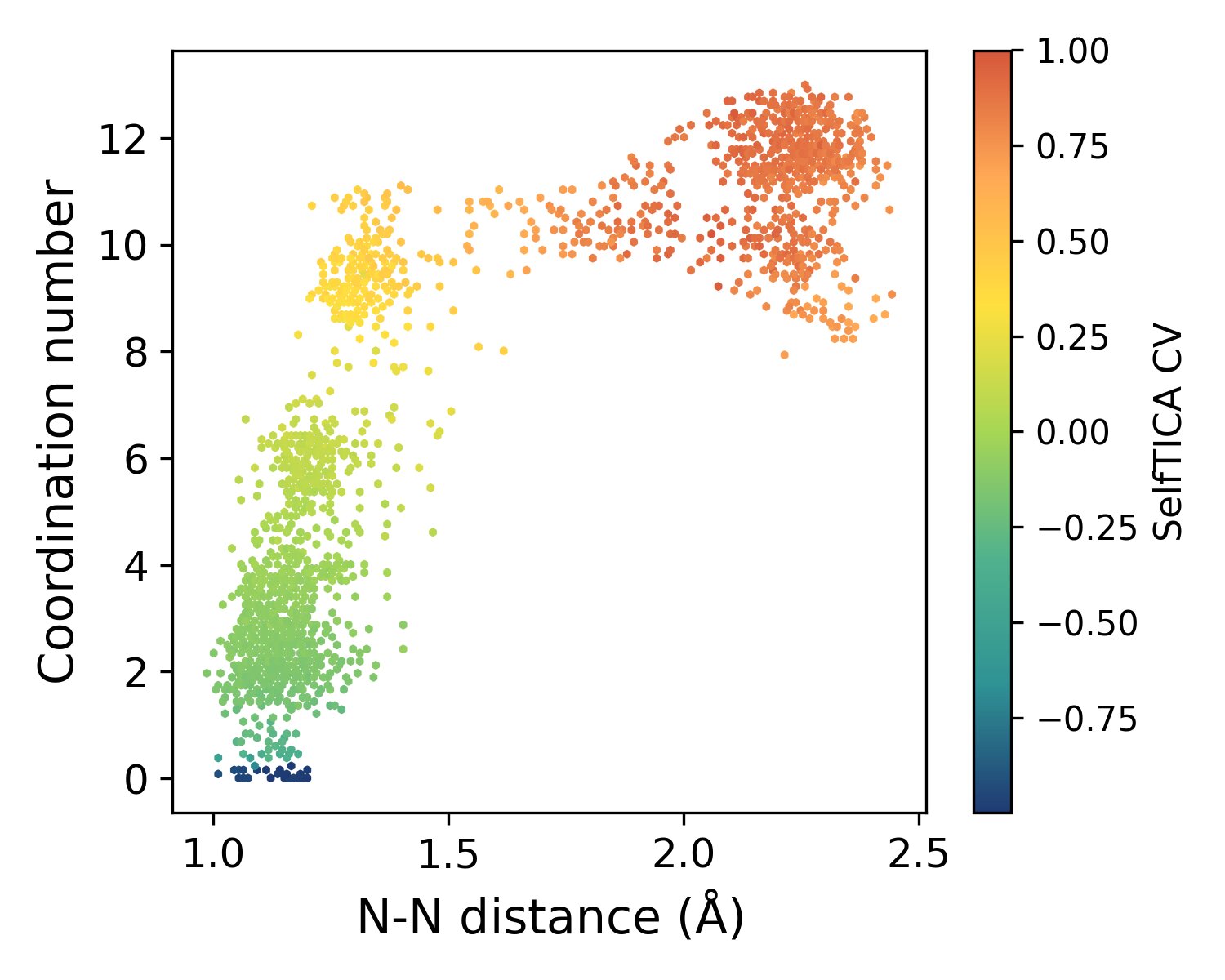}
\caption{Projection of the SelfTICA CV onto the two-dimensional space defined by the $\mathrm{N\!-\!N}$ distance and the coordination number.}
\label{fig:fen2_cv}
\end{figure}

\vspace{-0.5cm}

\begin{figure}[H]
\centering
\includegraphics[width=1.0\linewidth]{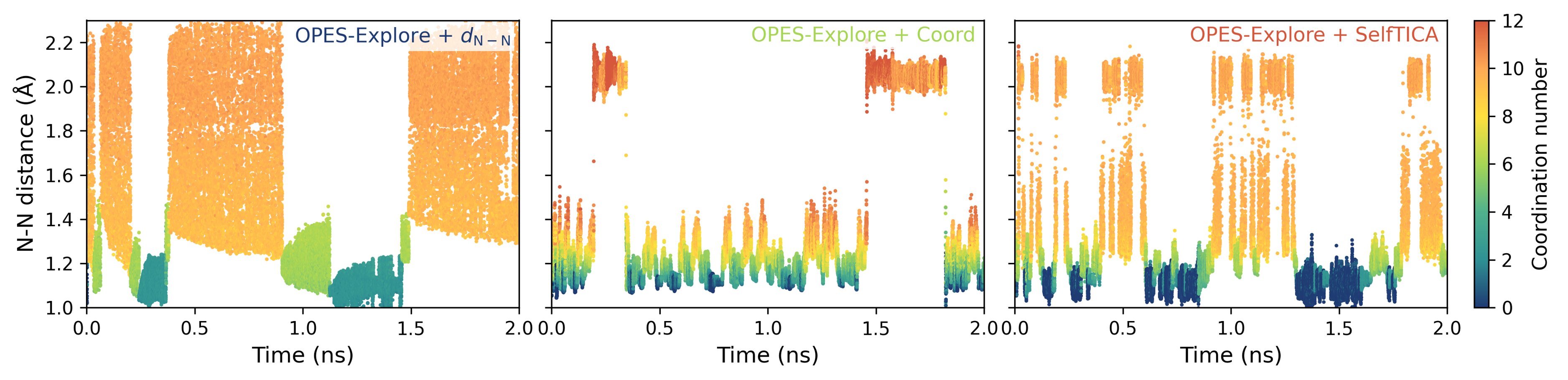}
\caption{Time evolution of the $\mathrm{N\!-\!N}$ distance from three independent OPES-explore simulations driven by the $\mathrm{N\!-\!N}$ distance, coordination number, and SelfTICA CV, respectively. Data points are colored according to the coordination number.}
\label{fig:fen2_explore}
\end{figure}

\clearpage

\section{Transferring pretrained SelfTICA representations to committor learning}

\subsection{Computational details}

\paragraphtitle{Committor model training details}
For committor learning, the pretrained SelfTICA encoder was kept fixed throughout training, and only a lightweight task-specific head was optimized to predict the committor $q_\theta(\mathbf{R})$. The training objective combined a boundary loss, which enforces $q_\theta=0$ and $q_\theta=1$ for configurations in states $A$ and $B$, respectively, with the variational Kolmogorov objective used to optimize the committor outside the boundary states. The relative weight of the boundary loss was set to $\alpha=100$ for the triple-well potential, $\alpha=10$ for alanine dipeptide, and $\alpha=1$ for chignolin. For all systems, the task-specific head was optimized using Adam with an initial learning rate of $1\times10^{-3}$ and a weight decay of $1\times10^{-5}$. An exponential learning-rate scheduler with a decay factor of $\gamma=0.99999$ was applied throughout training. The resulting committor was subsequently used to construct the Kolmogorov bias and drive OPES-explore sampling toward transition-relevant configurations, with $\lambda$ controlling the strength of the Kolmogorov bias.

For the triple-well potential, the committor head was trained in a single transfer round using boundary-state trajectories of 2,000 time units from each of states \(A\) and \(B\), together with an unbiased trajectory of 10,000 time units. The downstream head was optimized for 8,000 epochs while the pretrained SelfTICA encoder remained frozen. The resulting committor was then used to drive Kolmogorov-biased OPES sampling with \(\lambda=1.0\) and an OPES barrier of \(10\,k_{\mathrm B}T\) for a total of 10,000 time units. The final \(p_{\mathrm K}\) distribution was evaluated from an independent committor-biased production trajectory.

For alanine dipeptide, committor learning was performed in two successive transfer rounds while keeping the pretrained SelfTICA encoder fixed. In Transfer 0, the downstream head was trained for 5,000 epochs using 10~ns of boundary-state data from states $A$ and $B$. The resulting committor was then used to drive Kolmogorov-biased OPES sampling for 20~ns to generate transition-region configurations. In Transfer 1, data collected from the Transfer 0 simulation were combined with the original boundary-state data, and the downstream head was refined for an additional 5,000 epochs while the SelfTICA encoder remained frozen. In both rounds, $\lambda$ was set to 0.8 and the OPES barrier to $30$~kJ~mol$^{-1}$, with each biased simulation performed for 20~ns. The final $p_{\mathrm K}$ distribution was evaluated from an independent 20~ns committor-biased production trajectory.

For chignolin, committor learning was similarly performed in two successive transfer rounds while keeping the pretrained SelfTICA encoder fixed. In Transfer 0, a lightweight nonlinear readout consisting of a single hidden layer with 16 units and a hyperbolic-tangent activation was trained for 100,000 epochs using 100~ns of boundary-state data from the folded and unfolded ensembles on top of the frozen five-dimensional SelfTICA representation. The resulting committor was then used to drive Kolmogorov-biased OPES sampling for 250~ns. In Transfer 1, data collected from two Transfer 0 biased simulations were combined with the original boundary-state data, and the task-specific head was refined for 10,000 epochs while the pretrained SelfTICA representation remained unchanged. In both rounds, $\lambda$ was set to 1.0 and the OPES barrier to $20$~kJ~mol$^{-1}$, with each biased simulation performed for 250~ns. The final $p_{\mathrm K}$ distribution was evaluated from an independent 25~ns committor-biased production trajectory.

\subsection{Additional results}

\begin{figure}[H]
\centering
\includegraphics[width=0.4\linewidth]{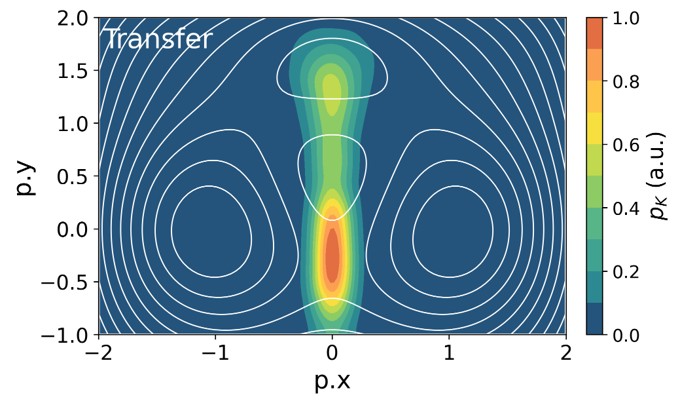}
\caption{Kolmogorov distribution obtained from the transferred committor for the triple-well potential.The committor was trained using boundary configurations from states $A$ and $B$ together with configurations from an unbiased trajectory.}
\label{fig:transfer_tri-well_pk}
\end{figure}

\begin{figure}[H]
\centering
\includegraphics[width=0.8\linewidth]{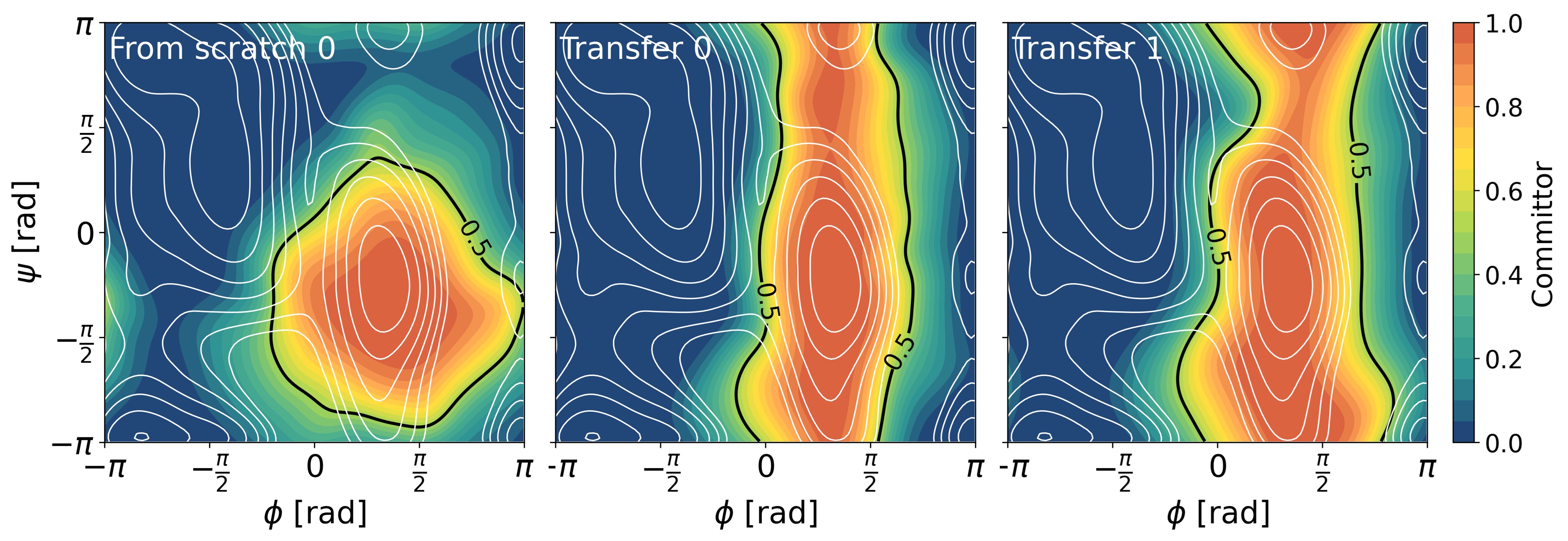}
\caption{Transfer and refinement of the committor for alanine dipeptide. Transfer 0 was trained using boundary configurations only, whereas Transfer 1 additionally incorporated transition-region configurations generated from the Transfer 0 simulation.}
\label{fig:transfer_ala2_committor}
\end{figure}

\begin{figure}[H]
\centering
\includegraphics[width=0.4\linewidth]{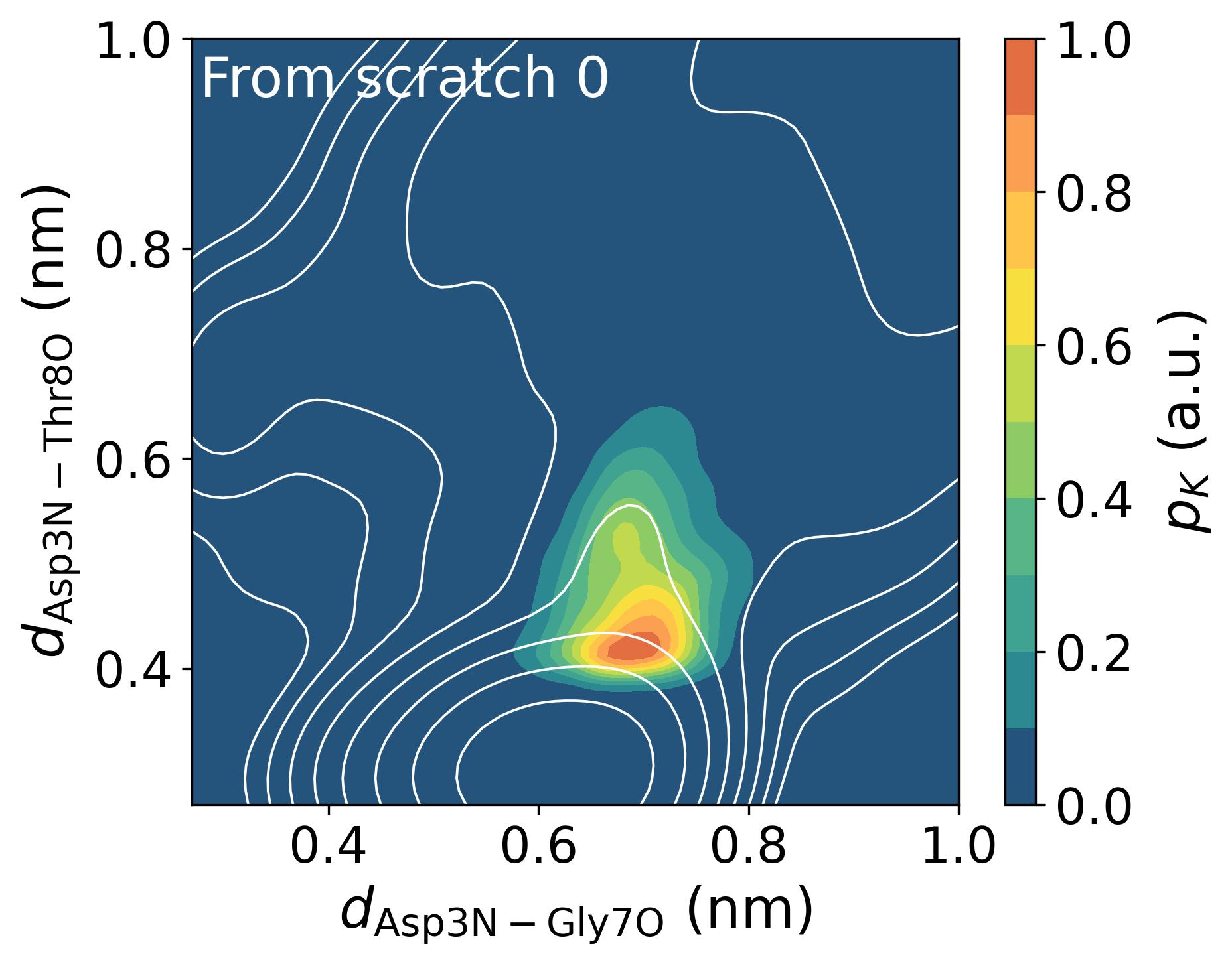}
\caption{\justifying
Kolmogorov distribution, $p_K$, from the initial committor-guided sampling of chignolin using a model trained from scratch on boundary and unbiased configurations, projected onto the Asp3N--Gly7O and Asp3N--Thr8O distances.}
\label{fig:transfer_tri-well_pk}
\end{figure}

\end{onecolumngrid}

\end{document}